\documentclass[11pt,article,onecolumn]{IEEEtran} 
\usepackage{cite}
\usepackage{epsf}
\usepackage{subfigure}

\newtheorem{lemma}{Lemma}

\def\vec#1{{\bf #1}}

\def\blambda{\mbox{\boldmath$\lambda$}}
\def\bgamma{\mbox{\boldmath$\gamma$}}

\def\bmu{\mbox{\boldmath$\mu$}}
\def\Tr#1{\mbox{Tr}\left\{ #1 \right\} }
\def\sgn{\mbox{sgn}}

\def\bm#1{{\mbox{\boldmath{$#1$}}}}

\def\nt{{n_t}}
\def\nr{{n_r}}

\def\bG{{\bf G}}

\def\bU{{\bf U}}
\def\bV{{\bf V}}

\begin{document}
\bstctlcite{BSTcontrol}

\title{Capacity and Character Expansions:
Moment generating function and other exact results
for MIMO correlated channels} 

\author{Steven H. Simon, Aris L. Moustakas and Luca Marinelli%
\thanks{Submitted to IEEE Transactions on Information Theory, March 2004}
\thanks{S. H. Simon (email: shsimon@bell-labs.com), A. L. Moustakas
(email: arislm@science.bell-labs.com), and L. Marinelli are with
Bell Labs, Lucent Technologies.}}



\maketitle

\begin{abstract}
We apply a promising new method from the field of representations
of Lie groups to calculate integrals over unitary groups, which
are important for multi-antenna communications. To demonstrate the
power and simplicity of this technique, we first re-derive a
number of results that have been used recently in the community of
wireless information theory, using only a few simple steps. In
particular, we derive the joint probability distribution of
eigenvalues of the matrix $\bG\bG^\dagger$, with $\bG$ a
semicorrelated Gaussian random matrix or a Gaussian random matrix
with a non-zero mean. These joint probability distribution
functions can then be used to calculate the moment generating
function of the mutual information for Gaussian MIMO channels with
these probability distribution of their channel matrices $\bf G$.
We then turn to the previously unsolved problem of calculating the
moment generating function of the mutual information of MIMO
channels, which are correlated at both the receiver and the
transmitter. From this moment generating function we obtain the
ergodic average of the mutual information and study the outage
probability. These methods can be applied to a number of other
problems. As a particular example, we examine unitary encoded
space-time transmission of MIMO systems and we derive the received
signal distribution when the channel matrix is correlated at the
transmitter end.
\end{abstract}

\begin{keywords}
Capacity, Space-Time Coding, Multiple Antennas, Unitary Space-Time
coding, MIMO, Random Matrix Theory, Character expansions.
\end{keywords}

\section{Introduction}
\label{Introduction}

\PARstart{I}{n} recent years, a flurry of research has resulted
from the prediction\cite{Foschini1998_BLAST1, Telatar1995_BLAST1}
that the use of multiple antennas to transmit and receive signals
leads to substantial increases in information throughput. In an
effort to quantify these throughput gains, several authors have
analyzed the asymptotic behavior of the
multiple-input-multiple-output (MIMO) mutual information for large
antenna numbers. In \cite{Verdu1999_MIMO1,
Rapajic2000_InfoCapacityOfARandomSignatureMIMOChannel} the
asymptotic  ergodic capacity (mutual information averaged over
channel realizations) was evaluated in closed form for Gaussian
i.i.d. channels. In \cite{Moustakas2000_BLAST1, Chuah2002_MIMO1,
Muller2002_RandomMatrixMIMO, Tulino2004_INDLargeNCapacity} the
assumption of i.i.d. channels was lifted and the effects of
spatial correlations were included. In addition, asymptotic
results on the variance of the mutual information were derived,
\cite{Moustakas2003_MIMO1, Sengupta2000_BLAST1,
Hochwald2002_MultiAntennaChannelHardening,
Biglieri2003_LargeSystemAnalysisOfMIMOCapacities}, which together
with its mean were shown \cite{Moustakas2003_MIMO1} to describe
the full distribution very accurately even for small antenna
numbers.

More recently, several exact results regarding the distribution of
mutual information for the MIMO link have appeared in the
literature. Using different methods these prior works calculate
the moment generating function of the distribution of the mutual
information for various assumptions about the statistics of the
channel matrix. From the generating function, the moments can be
generated by direct differentiation, or the probability of outage
\cite{Ozarow1994_OutageCapacity} can be derived through a simple
numerical integral. Specifically,
 \cite{Wang2002_OutageMutualInfoOfSTMIMOChannels} first extracted
 the moment-generating function for Gaussian uncorrelated
channels. Furthermore, in
\cite{Kang2002_CapacityOfMIMORicianChannels,
Kang2003_CapacityOfMIMORicianPlusInterferers} Rician channels were
used. In addition,
\cite{Chiani2003_CapacityOfSpatiallyCorrelatedMIMOChannels,
Smith2003_CapacityMIMOsystems_with_semicorrelated_flat_fading}
treated the case of semi-correlated channels, where either the
transmitting or receiving antennas are correlated (but not both).
In particular,
\cite{Chiani2003_CapacityOfSpatiallyCorrelatedMIMOChannels} dealt
with the case where the number of correlated antennas is less than
or equal to the number of uncorrelated antennas, while
\cite{Smith2003_CapacityMIMOsystems_with_semicorrelated_flat_fading}
dealt with the opposite case. All above progress in the
calculation of the moment generating function relied essentially
on previously known results in the theory of Wishart matrices
(matrices of the form $\bG \bG^\dagger$), which allowed the
calculation of the joint probability distribution of eigenvalues
of such matrices.

Interestingly, one can formulate the problem of calculating the
joint probability distribution of eigenvalues of $\bG\bG^\dagger$
in terms of integrals over the unitary group. Such integrals have
been analyzed since the 1980's in the field of high energy
physics\cite{ItzyksonZuber1980_PlanarApproximation,
Brower1981_LatticeQCD} and even earlier in the mathematics
community\cite{HarishChandra1957_DifferentialOperatorsSemiSimpleLieAlgebras}.
More recently, \cite{Balantekin2000_CharacterExpansionsETC} recast
the same problem from an elegant group-theoretic point of view,
which allows one to perform such integrals in a few simple steps.
In this paper we will apply these new group-theoretic methods to
calculate the joint distribution of eigenvalues of
$\bG\bG^\dagger$, for a number of different channel distributions,
$p(\bG)$. From the eigenvalue distribution the moment generating
function of the mutual information is readily obtained. We first
simply re-derive a few of the above mentioned prior results
\cite{Chiani2003_CapacityOfSpatiallyCorrelatedMIMOChannels,
Smith2003_CapacityMIMOsystems_with_semicorrelated_flat_fading,
Kang2002_CapacityOfMIMORicianChannels} and then we obtain a new
result of the moment generating function for zero-mean Gaussian
channels with correlations in both (transmitting and receiving)
ends of the link.\footnote{In
\cite{Simon2004_EigenvalueDensityOfCorrRandomWishartMatrices} we
derive a similar result used to calculate the marginal eigenvalue
distribution of $\bG\bG^\dagger$ with such statistics.} In
addition, we discuss in an appendix how similar techniques can be
powerful in analyzing unitary space-time encoded transmission in
MIMO systems.\cite{Marzetta1999_USTM, Hochwald2000_USTM,
Hughes2000_DUSTM, Hassibi2002_USTM}.

 The techniques we use are both simple and
powerful, and a main aim of this paper is to expose the reader to
these methods. Although we cannot attempt to give a complete
exposition of the field of group-theory, we are able to give a
small and simple set of rules, definitions, and theorems that will
allow the interested reader to apply these techniques to other
problems of interest.

\subsection{Outline}
\label{sec:Outline}

We start with some notational definitions and the introduction of
the system model in Section \ref{sec:Definitions}. In Section
\ref{sec:Useful Results} we introduce the reader to some basic
definitions in group theory and quote the basic results of the
method, which will be applied in the remainder of the paper. We
illustrate the main steps of this method by calculating the known
joint probability distribution of the eigenvalues of
$\bG\bG^\dagger$ for the semicorrelated Gaussian MIMO channel
$\bG$ in Section \ref{sec:Case of semi-correlated G with zero
mean}, which involves a single unitary integral. In Section
\ref{sec:Case of uncorrelated G with non-zero mean G0} we
generalize our method to derive the joint distribution of
eigenvalues of $\bG\bG^\dagger$ for $\bG$ being a non-zero mean
Gaussian MIMO channel. Although also a known result, this analysis
gives us further insight in the utility of our method, since this
case requires a double unitary integration.

In Sections \ref{sec:Case of fully correlated G with zero mean}
and \ref{sec:Moment-Generating Function with no Interference} we
apply the technique to the case the zero-mean Gaussian channel
with correlated transmitting and receiving antennas and derive a
closed-form expression for the moment generating function for the
corresponding mutual information, which is a new result. We use
this formula in Section \ref{sec:Ergodic Capacity for fully
correlated channels} to obtain an analytical expression for the
ergodic capacity and also study the outage probability.

As mentioned above, in this paper we aim to expose the reader to
the new group-theoretic methods discussed in Section
\ref{sec:Useful Results}. These methods are also applicable in
other areas of information theory of MIMO systems. To illustrate
this point, we apply this method in the context of unitary
space-time coding and calculate the received signal distribution
for the case where the channel is correlated at the transmitter
end, a new result. In order not to disrupt the flow of the main
part, this analysis appears in Appendix \ref{app:Application of
Unitary Integrals on Unitary Space-Time Encoding}.

\section{Definitions} \label{sec:Definitions}

\subsection{Notation}
\label{sec:Notation}

\subsubsection{Vectors/Matrices}\label{sec:Vectors/Matrices}
Throughout this paper we will use bold-faced upper-case letters to
denote matrices, e.g. $\vec X$, with elements given by $X_{ab}$.
Bold-faced lower-case letters, e.g. $\vec x$ will be used for
vectors with elements $x_i$ and non-bold lower-case letters will
denote scalar quantities. Also, the superscripts $T$ and $\dagger$
will indicate transpose and Hermitian conjugate operations and
$\vec I_n$ will represent the $n$-dimensional identity matrix.
Finally, $diag(\vec x)$ will denote a diagonal, but not
necessarily square matrix with the elements of the vector $\vec x$
on the main diagonal.

\subsubsection{Determinants}\label{sec:Determinants}
$\det(f(i,j))$ will denote the determinant of a matrix with the
$i,j$th element given by $f(i,j)$, an arbitrary function of $i$
and $j$. In order to keep the notation consistent, we will always
use the indices $i,j$ for this purpose in determinants.
Furthermore, we will be using the notation $\Delta(\vec x)$ for an
$n$-dimensional Vandermonde determinant of the elements of the
$n$-dimensional vector $\vec x$.
\begin{equation}\label{eq:Vandermonde_det_def}
  \Delta(\vec x) = \Delta([x_1,\ldots,x_n]) = \det(x_i^{j-1}) = \prod_{i>j} (x_i-x_j)
\end{equation}
In addition, due to the extensive use of determinant expansion in
the text, it is convenient to define the following notation
\begin{equation}\label{eq:det_expansion_def}
  \det\vec X = \sum_{\vec a} \sgn(\vec a) \prod_{i=1}^M
  X_{ia_i} = \frac{1}{M!} \sum_{\vec a}\sum_{\vec b}
   \sgn(\vec a)\,  \sgn(\vec b) \prod_{i=1}^M X_{a_ib_i}
\end{equation}
where $\vec X$ is a $M$-dimensional matrix. In the above
expression the vector $ \vec a = [a_1, a_2,\ldots, a_M]$ is a
permutation of the sequence $[1,2,\ldots, M]$. $\sgn(\vec a)$ is
the sign of the permutation $a$, which is $\sgn(\vec a)=+1$ if
$[a_1, a_2,\ldots, a_M]$ is an even permutation of the sequence
$[1,2,\ldots, M]$, $\sgn(\vec a)=-1$ for an odd permutation. It is
implied here that if $\vec a$ is not a permutation of
$[1,2,\ldots,M]$, $\sgn(\vec a)=0$. The summation is over $1\leq
a_i\leq M$ for $i=1,\ldots,M$.

\subsubsection{Integral Measures}\label{sec:Integral Measures}
We will use three types of integrals: In the first we will be
integrating over the real and imaginary part of the elements of a
complex $m_{rows}\times m_{cols}$ matrix $\vec X$. The integral
measure will be denoted by
\begin{equation}
\label{eq:DX} %
{\cal D}\vec X = \prod_{a=1}^{m_{rows}}\prod_{\alpha=1}^{m_{cols}}
\frac{d{\rm Re}X_{a\alpha}\,d{\rm Im}X_{a\alpha}}{\pi}
\end{equation}
We will also be using the notation \begin{equation} d\vec
x=\prod_{i=1}^n d x_i \end{equation} for integrals over the
positive real axis of the elements of the vector $\vec x$.
Finally, $D\vec U$ will represent the Haar integral measure
\cite{Sternberg_GroupTheory_book} for integrals over the elements
of the unitary group $U(M)$ for a specified $M$.

\subsubsection{Antenna Numbers}\label{sec:Antenna Numbers}
$\nt$ and $\nr$ will denote the number of antennas at the
transmitter and receiver, respectively. We will also use the
following notation for maxima and minima of antennas:
$M=\max(\nt,\nr), N=\min(\nr,\nt)$.

\subsection{System Model}
\label{sec:System Model}

We consider the case of single-user transmission from $\nt$
transmit antennas to $\nr$ receive antennas over a narrow band
fading channel. The received $\nr$-dimensional complex signal
vector $\vec y$ can be written as
\begin{equation}\label{eq:basic_channel_eq}
  \vec y =  \bG \vec x + \vec z
\end{equation}
$\bG$ is a $\nr\times\nt$ complex matrix of the channel
coefficients from the transmitting to the receiving arrays. $\vec
x$ is the $\nt$-dimensional vector of transmitted Gaussian
signals, while $\vec z$ is the $\nr$-dimensional additive Gaussian
noise vector. Without loss of generality, both $\vec x$ and $\vec
z$ are taken to be i.i.d., zero-mean with unit-variance. One can
straightforwardly include arbitrary covariances for both in the
end result (e.g. see \cite{Moustakas2003_MIMO1}). Finally, the
channel $\bG$ is assumed to be block-fading, i.e. constant over a
significant amount of time and then changing completely according
to its statistics. For concreteness, Gaussian channel statistics
of $\bG$ are assumed.

During reception, the receiver is assumed to know the
instantaneous channel matrix $\bG$.  Thus, the mutual information
can be expressed as \cite{Foschini1998_BLAST1}
\begin{equation}\label{eq:cap_def}
I\left(\vec y ; \vec x\left|\bG\right.\right) = \log\det\left(\vec
I_\nt +  \bG {}^\dagger \bG\right)
\end{equation}
The $\log$ above (and throughout the whole paper)  represents the
natural logarithm and thus $I$ is expressed in nats.

We describe the statistics of the channel $\bG$ in terms of its
probability distribution $p(\bG)$. Due to the underlying
randomness of $\bG$, the mutual information $I$ is also a random
quantity. In this paper we will analyze the statistics of the
mutual information $I$ in (\ref{eq:cap_def}), assuming that the
channel $\bG$ is a Gaussian random matrix.  We will specify
details of several different types of Gaussian distributions to
study below (iid, semicorrelated, nonzero mean, and correlated).
In order to analyze the statistics of the mutual information $I$,
we will calculate the generating function $g(z)$ defined as,
\begin{equation}\label{eq:g_G_def}
  g(z)  = E_{\bG}\left[ e^{z I}  \right] = E_{\bG}\left[
  \left\{ \det\left(\vec I_\nt +  \bG {}^\dagger \bG\right) \right\}^{z}\right]
\end{equation}
where the expectation $E_{\bG}[\,\cdot\,]$ is with respect to the
probability distribution $p(\bG)$ of the matrix $\bG$. Once $g(z)$
is known, all moments of $I$ can be generated by evaluating the
derivatives of $g(z)$ at $z=0$.
\begin{eqnarray}\label{eq:g_z_derivative1}
  \left. g'(z)\right|_{z=0} &=& E_{\bG}\left[I\right] \\
  \left. g''(z)\right|_{z=0} &=&  E_{\bG}\left[I^2\right]
\label{eq:g_z_derivative2}
\end{eqnarray} etc. In addition, the
probability of outage, $P_{out}$, i.e. the probability that the
mutual information is less than a given value $I_{out}$
\cite{Ozarow1994_OutageCapacity}, can be readily obtained by
performing the integral
\begin{equation}\label{eq:outage_integral_def}
  P_{out} =  E_{\bG}\left[\Theta\left(I-I_{out}\right)\right] = \int_{-\infty}^{+\infty}\frac{dz}{2\pi i} g(iz)
  \frac{e^{-izI_{out}}}{z-i0^+}
\end{equation}
where we have used the identity $\Theta(x) = \int dz/(2\pi i)
\exp(ixz)/(z-i0^+)$ for the Heavyside step function $\Theta(x)$.

{}From the form of the mutual information in (\ref{eq:cap_def})
and (\ref{eq:g_G_def}), it is clear that the generating function
can be written very simply in terms of the eigenvalues $
\lambda_i$ of the matrix $\bG^\dagger\bG$, as
\begin{equation}\label{eq:g_z_lambda}
  g(z) = E_{\blambda}\left[\prod_{i=1}^N(1+\lambda_i)^z\right] =
  \prod_{i=1}^N \int_0^\infty d\lambda_i (1+\lambda_i)^z P(\{\lambda_j\})
\end{equation}
where the product is over the $N=\min(\nt,\nr)$ non-zero
eigenvalues of $\bG^\dagger\bG$ and the expectation
$E_{\blambda}[\,\cdot\,]$ is defined as a multiple integral over
the joint distribution $P(\{\lambda_i\})$ of the eigenvalues
$\{\lambda_i\}$. Note that $P(\{\lambda_i\})$ will be different
for each distribution $p(\vec G)$ we will analyze.

\section{Important Results from Representation Theory of Unitary
Groups}\label{sec:Useful Results}

In this section we will briefly introduce a few relevant
quantities and a number of important results from the theory of
representations of unitary groups, which will be used in the
paper. The interested reader can refer to several textbooks,
including \cite{Sternberg_GroupTheory_book, Hua_book_GroupTheory}.
We do not attempt a thorough exposition of the subject of group
theory. However, we do aim to give a self-contained introduction,
so that the interested reader can effectively apply these methods
to tackle other problems of interest.

A $d$-dimensional representation $V$ of a group $G$ is a
homomorphism from $G$ into a group of $d$-dimensional invertible
matrices. Such groups include $Gl(M)$, the group of complex
invertible matrices of dimension $M$, and $U(M)$ is its subgroup
of unitary matrices (i.e., the subgroup of matrices $\bU$ such
that $\bU \bU^\dagger = \bf I$). These groups are also called Lie
groups, i.e. differentiable manifolds with both multiplication and
inverse operations being differentiable.

\subsubsection{Irreducible Representations}
\label{def: Irreducible Representations}

A $d$-dimensional representation is {\em irreducible} if it has no
non-trivial invariant subspaces. In other words, a representation
is irreducible if there exists no $d$-dimensional invertible
matrix $\bf A$,  such that $\bf A H A^{-1}$ becomes block diagonal
for all group elements $\bf H$.

The irreducible representations of the unitary group $U(M)$ (and
$Gl(M)$) can both be labelled by an $M$-dimensional vector $\vec
m=[m_1,m_2,\ldots,m_M]$, with integers $m_1\geq m_2\geq\ldots \geq
m_M\geq 0$.\cite{Sternberg_GroupTheory_book, Hua_book_GroupTheory}

\subsubsection{Dimension of Irreducible Representations}
\label{def:Dimension of Irreducible Representations} The dimension
$d_{\vec m}$ of an irreducible representation labelled by $\vec m$
is the dimension of its invariant subspace. For both cases of
$U(M)$ and $Gl(M)$ it can be shown
\cite{Balantekin1981_DimensionAndCharacterFormulasForLieSupergroups}
that $d_{\vec m}$ is given by
\begin{equation}\label{eq:d_r_definition}
  d_{\vec m} = \left[\prod_{i=1}^M \frac{(M+m_i-i)!}{(M-i)!}\right] \det\left[\frac{1}{(m_i-i+j)!}\right]
\end{equation}
where the matrix elements inside the determinant with $m_i-i+j<0$
are zero.

The above form of the dimension $d_{\vec m}$ in terms of a
determinant of a matrix of inverse factorials is not particularly
useful for manipulation. The following lemma makes its form more
manageable. Its proof appears in Appendix \ref{app:Proof of Lemma
Vandermonde determinant form of d_r}.

\begin{lemma}[Vandermonde determinant form of $d_{\vec m}$]
\label{lemma:Vandermonde determinant form of d_r}
 The dimension
$d_{\vec m}$ of an irreducible representation $\vec m$ of $U(M)$,
given in (\ref{eq:d_r_definition}) can be written as
\begin{equation}\label{eq:d_r_vandermonde_form}
  d_{\vec m} = \left[\prod_{i=1}^M \frac{1}{(M-i)!}\right] \left(-1\right)^{\frac{M(M-1)}{2}}
  \Delta(\vec k)
\end{equation}
where $\Delta(\cdot)$ represents the Vandermonde determinant and
the vector $\vec k$ has elements
\begin{equation}\label{eq:vector_k_def_app}
  k_i = m_i -i + M
\end{equation}
for $i=1,\ldots,M$ and $m_i$ are the elements of the
representation vector $\vec m$.
\end{lemma}

\subsubsection{Character of Representation} \label{def: Characters
of Representations} The character $\chi(g)$ of a group element $g$
in the representation $V$ is equal to the trace of the
corresponding matrix, i.e. $\chi(g) = \Tr{V(g)}$. Also, $\chi(g)$
depends only on the eigenvalues of $V(g)$. Calculating the
characters of irreducible representations is greatly facilitated
by Weyl's character formula \cite{Weyl_book,
Hua_book_GroupTheory}, which for  $U(M)$ and $Gl(M)$ takes the
form:
\begin{equation}\label{eq:Weyl_formula_def}
\chi_{\vec m}(\vec A) = \Tr{\vec A^{(\vec m)}} =
\frac{\det\left(a_i^{m_j+M-j}\right)}{\Delta(a_1,\ldots,a_M)}
\end{equation}
where the index $\vec m$ denotes the irreducible representation
 $(m_1,\ldots,m_M)$ and $a_i$, for
$i=1,\ldots,M$, are the eigenvalues of $\vec A$ in the fundamental
($M$-dimensional) representation. Here $\vec A$ in $\chi_{\vec
m}(\vec A)$ represents a $M$-dimensional matrix corresponding to
the group element that is being represented, while $\vec A^{(\vec
m )}$ represents the $d_{\vec m}$-dimensional matrix, which is the
$(\vec m)$-representation of $\vec A$.

For example, the characters of the one-dimensional unitary group
$U(1)$ are given by $\chi_n(e^{i\phi})=e^{in\phi}$, where the
character index $n$ takes non-negative integer values and
$e^{i\phi}$ is an arbitrary element in $U(1)$. Thus a Fourier
transform\footnote{The astute reader will realize that the
characters have only nonnegative $n$, whereas the Fourier
transform has all $n$.  This means the Fourier transform is
actually a character expansion of $U$ and $U^\dagger$.  This
complication does not enter into our particular use of the
character expansion.} can be seen as an expansion in the
characters of $U(1)$. This suggests that the characters of a group
can form a good basis for expanding functions, which are invariant
under the corresponding group operations. Although such expansions
are generally complicated, recently
\cite{Balantekin2000_CharacterExpansionsETC} showed how it can be
done for exponentials of a trace of a $Gl(M)$ matrix. His result
can be summarized in the following important lemma.

\begin{lemma}[Character Expansion of Exponential]
\label{lemma:Character_Expansion} Let $\vec A$ be a member of the
general linear group $Gl(M)$ and $x$ an arbitrary constant. Then
the following equation holds:
\begin{equation}\label{eq:character_expansion}
e^{x\Tr{\vec A}} = \sum_{\vec m} \alpha_{\vec m}(x) \chi_{\vec
m}(\vec A)
\end{equation}
In the above expression, $\chi_{\vec m}(\vec A)$ is the character
of $\vec A$ in the representation $\vec m$ and the sum is over all
irreducible representations of $Gl(M)$ parameterized with the
vector $\vec m=[m_1,m_2,\ldots,m_M]$, with integers $m_1\geq
m_2\geq\ldots \geq m_M\geq 0$. In (\ref{eq:character_expansion})
the coefficient for each character, $\alpha_{\vec m}(x)$ is given
by
\begin{equation}\label{eq:alpha_x_definition}
  \alpha_{\vec m}(x)= \left(x\right)^{m_1+m_2+\ldots+m_M}
  \left[\prod_{i=1}^M \frac{(M-i)!}{(M+m_i-i)!}\right] d_{\vec m}
\end{equation}
where $d_{\vec m}$ is the dimension of the irreducible
representation and is given by (\ref{eq:d_r_definition}).
\end{lemma}

The importance of this result is that it allows us to expand the
exponential in group-invariant quantities as the characters
$\chi_{\vec m}$.

\subsubsection{Orthogonality Relation between Unitary Group Matrix
Elements}\label{def:integration_over_U_U_bar} %
The integration properties of elements of the unitary group in
different representations are given in the following (see
\cite{Sternberg_GroupTheory_book})
\begin{equation}
\label{eq:def_integration_over_U_U_bar}
  \int D{\bf U}\,\, U^{(\vec m)}_{ij} U^{(\vec
  m')*}_{kl}=\frac{1}{d_{\vec m} }\delta_{\vec m \vec m'}
  \delta_{ik}\delta_{jl}
\end{equation}
where ${\vec m}$ is the dimension of the representation and $D\bU$
is the standard Haar integration measure over $U(M)$.

\subsubsection{Cauchy-Binet Formula}
Finally, we cite a useful result, which will be applied in our
calculations below (see proof in \cite{Hua_book_GroupTheory}).
\begin{lemma}[Cauchy-Binet Formula]
\label{lemma:Cauchy-Binet Formula} Given $M$-dimensional vectors
$\vec a$ and $\vec b$, and a function $W(z) = \sum_{i=0}^\infty
w(i) z^i$ convergent for $|z| < \rho$, then, if $|a_i b_j| < \rho$
for all $i,j$, we have:
\begin{equation}
\label{eq:cauchy} \sum_{k_1 > k_2\ldots > k_m\geq 0} \!\!\!\!\!
\det[a_i^{k_j}]
    \det[b_i^{k_j}]\, \mbox{$\prod_{i=1}^m$} w(k_i) = \det[W(a_i b_j)]
\end{equation}
where the determinants are all taken with respect to the indices
$i$ and $j$.
\end{lemma}

These few definitions and formulae will be enough to allow us to
manipulate integrals over the unitary group in extremely powerful
ways.

\section{Calculation of the Joint Distribution of Eigenvalues
of $\bG\bG^\dagger$ using the Character Expansion}
\label{sec:Calculation of Joint Distribution of Eigenvalues}

As we saw in the end of Section \ref{sec:System Model}, one can
calculate $g(z)$ by first calculating the joint probability
density  $P(\{\lambda_i\})$ of the eigenvalues $\lambda_i$. In
this section we will introduce a new method to calculate the joint
probability density for Gaussian channel matrices $\bG$ with
non-trivial correlations or non-zero mean. As we shall see, this
method allows us to obtain $P(\{\lambda_i\})$ with minimal effort,
using the results introduced in Section \ref{sec:Useful Results}.
Given its power and simplicity, we will expose the basic steps of
the approach in the main part of the paper, even though the
results presented in this section have been derived elsewhere,
albeit using more involved and less applicable methods.

To express the probability density in terms of the eigenvalues
$\lambda_i$ of $\bG\bG^\dagger$, it is convenient to change
integration variables, from the elements of $\bG$ to its singular
values,
\begin{equation}\label{eq:svd_def}
\mu_i=\sqrt{\lambda_i} = \mbox{SVD}(\bG)
\end{equation}
for $i=1,\ldots,M$, using the standard transformation
\begin{equation}\label{eq:GeqUmuV_def}
  \bG = \bU  diag(\bmu) \bV^\dagger
\end{equation}
where $\bU$, $\bV$ are unitary $\nt\times \nt$ and $\nr\times\nr$
matrices, respectively and $ diag(\bmu)$ is a diagonal
$\nt\times\nr$ matrix with diagonal elements $\mu_i$, for
$i=1,\ldots,N$. Including the Jacobian of the coordinate
transformation we get the following relation \cite{Mehta_book,
Edelman_thesis, Telatar1995_BLAST1,
Zheng2002_CommunicationOnGrassmannManifold}
\begin{equation}\label{eq:DG_eq_Jacobian_dl_dU_dV}
  {\cal D}\bG = {\cal C}_{M,N} \Delta(\blambda)^2 \, \prod_{i=1}^N
  \lambda_i^{M-N}
  d\blambda \, D\bU \, D\bV
\end{equation}
where $\blambda$ is an $N$-dimensional vector with elements
$\lambda_i$ and $\Delta(\blambda)$ is the corresponding
Vandermonde determinant. $d\blambda$ is the product of the
$\lambda_i$ differentials $d\blambda=d\lambda_1\ldots d\lambda_N$,
while $D\bU$ and $D\bV$ are the standard Haar integration measures
for the corresponding unitary matrices, see for example
\cite{Sternberg_GroupTheory_book}. The normalization constant
${\cal C}_{M,N}$ is given by the Laguerre-Selberg integral (see p.
354 in \cite{Mehta_book})
\begin{equation}\label{eq:C_MN_def}
  {\cal C}_{M,N}^{-1} = \prod_{j=1}^N
  \left[\Gamma(j+1)\Gamma(M-N+j)\right]
\end{equation}
Thus, we can express $P(\{\lambda_i\})$ as
\begin{equation}\label{eq:P_lambda_i_int1}
    P(  \{\lambda_i\}) = {\cal C}_{M,N} \Delta(\blambda)^2 \, \prod_{i=1}^N
  \lambda_i^{M-N} \int D\bU \int
    D\bV \,\,p(\bG=\bU diag(\bmu)\bV^\dagger)
\end{equation}
where the integrals are over all unitary matrices $\bU$, $\bV$ and
$p(\bG)$ is the probability distribution of the matrix $\bG$.

\subsection{Case of i.i.d. channel with zero mean}

When $\bG$ is independently and identically distributed (i.i.d.)
with zero mean, the probability distribution is given by
\begin{equation}
    p(\bG) = e^{-\Tr{\bG^\dagger \bG}}
\end{equation}
Substituting $\bG = \bU  diag(\bmu) \bV^\dagger$ we obtain
\begin{equation}
    p(\bG) = \exp\left[-\Tr{ diag(\bmu)^\dagger  diag(\bmu)}\right] = \exp\left[-\Tr{
    diag(\blambda})\right]
\end{equation}
which is independent of both $\bU$ and $\bV$. $diag(\blambda)$ is
an $N\times N$ diagonal matrix with diagonal elements
$\lambda_i=\mu_i^2$, which are the eigenvalues of
$\bG^\dagger\bG$. Thus, the integrals over the unitary group in
(\ref{eq:P_lambda_i_int1}) are trivial and we immediately obtain
the well-known result \cite{Mehta_book}
\begin{equation}\label{eq:P_lambda_i_iid}
    P(  \{\lambda_i\}) = {\cal C}_{M,N} \Delta(\blambda)^2 \prod_{i=1}^N
  \left[\lambda_i^{M-N} e^{-\lambda_i}\right]
\end{equation}
It is clear that this particular case is rather unique in that the
unitary integrals in (\ref{eq:P_lambda_i_int1}) are trivial. More
generally, the argument of these integrals will contain $\bU$ and
$\bV$ explicitly, making the joint distribution harder to obtain,
as we shall see below.

\subsection{Case of semi-correlated $\bG$ with zero mean}
\label{sec:Case of semi-correlated G with zero mean}

The simplest case of channel with a unitary matrix explicitly in
$p(\bG)$ is that of a semi-correlated channel, i.e. one with
non-trivial correlations on only one side of $\bG$, with the
probability distribution taking the form
\begin{equation}\label{eq:p_G_semicorr}
  p(\bG) = {\cal N}_{\vec T} \, \exp\left[-\Tr{{ \bG} {\vec  T}^{-1}{\bG}^\dagger}\right] = {\cal N}_{\vec T}
  \exp\left[-\Tr{diag(\tilde{\blambda})\bU^\dagger{\vec
  T}^{-1}{\bU}}\right]
\end{equation}
where the  normalization constant is ${\cal N}_{\vec T}^{-1} =
\det\vec T^\nr$ and $diag(\tilde{\blambda})$ is an
$\nt$-dimensional diagonal matrix with the first $N$ diagonal
elements $\tilde{\lambda}_i = \lambda_i$, for $i=1,\ldots,N$ and
the remaining $M-N$ ones (if $\nt=M>N=\nr$) taking for now
arbitrary values. In the end of the calculation we take the limit
$\tilde{\lambda}_{N+1},\ldots,\tilde{\lambda}_M\rightarrow 0$. As
a result, only $N$ non-zero eigenvalues of $\bG\bG^\dagger$ will
remain.

Inserting (\ref{eq:p_G_semicorr}) into (\ref{eq:P_lambda_i_int1}),
$P(\{\lambda_i\})$ can be expressed as
\begin{equation}\label{eq:P_lambda_i_semicorr_int}
    P(  \{\lambda_i\}) = {\cal C}_{M,N} {\cal N}_{\vec T} \,
    \Delta(\blambda)^2 \, \prod_{i=1}^N
  \lambda_i^{M-N}
    \int D\bU  \exp\left[-\Tr{diag(\tilde{\blambda})\bU^\dagger\vec
    T^{-1}\bU}\right]
\end{equation}

{\bf Step 1: Integration over $\vec U$}. %
We next need to integrate out the unitary matrix $\vec U$. Such
integrals are well known
\cite{HarishChandra1957_DifferentialOperatorsSemiSimpleLieAlgebras,
ItzyksonZuber1980_PlanarApproximation}, and we could easily skip
directly to the result (\ref{eq:P_lambda_i_semicorr_MN_gen}).
However, it is useful to go through this particular derivation, so
that the reader will become familiar with the technique. Applying
Lemma \ref{lemma:Character_Expansion} to the exponential of
(\ref{eq:P_lambda_i_semicorr_int}) we get
\begin{eqnarray}\label{eq:P_lambda_i_semicorr_char_exp}
    P(  \{\lambda_i\}) &=& {\cal C}_{M,N} {\cal N}_{\vec T}
    \,\Delta(\blambda)^2 \prod_{i=1}^N
    \lambda_i^{M-N}
   \int D\bU \, \sum_{\vec m} \alpha_{\vec m}(-1) \chi_{\vec m}\left( diag(\tilde{\blambda})  \bU^\dagger {\bf T}^{-1} \bU \right)
   \\ \nonumber %
 &=& {\cal C}_{M,N} {\cal N}_{\vec T}
 \,\Delta(\blambda)^2 \prod_{i=1}^N
    \lambda_i^{M-N}
 \int D\bU  \sum_{\vec m} \alpha_{\vec m}(-1) \Tr{ diag\left(\tilde{\blambda}\right)^{(\vec m)} \left(\bU^{(\vec
m)}\right)^\dagger \left({\bf T}^{(\vec m)}\right)^{-1} \bU^{(\vec
m)} }
\end{eqnarray}
where $\alpha_{\vec m}(x)$ is defined in
(\ref{eq:alpha_x_definition}) and the sum is over all integer
vectors $\vec m=[m_1,\ldots,m_\nt]$ with $m_1\geq m_2 \geq\ldots
\geq m_\nt\geq 0$, corresponding to the representations of
$U(\nt)$ (see (\ref{eq:character_expansion})). We remind the
reader that we use the notation $\bf M^{(m)}$ to be the group
element $\bf M$ in the representation $\bf m$. The second equation
here results from the definition of the character $\chi_{\vec m}$
as the trace of its argument. As mentioned above the integrand now
is {\em quadratic} in the unitary matrix elements and thus we may
integrate over the unitary matrix $\bU$ using the standard
orthogonality relation between unitary matrix elements, given in
Eq. (\ref{eq:def_integration_over_U_U_bar}). As a result we get
\begin{eqnarray} \label{eq:P_lambda_i_semicorr_char_exp_int_over_U}
P(  \{\lambda_i\}) &=& {\cal C}_{M,N} {\cal N}_{\vec T}
\,\Delta(\blambda)^2 \prod_{i=1}^N \lambda_i^{M-N} \sum_{\vec m}
\frac{\alpha_{\vec m}(-1)}{d_{\vec m}}
\Tr{\left({\bf T}^{(\vec m)}\right)^{-1}} \Tr{diag\left(\tilde{\blambda}\right)^{(\vec m)}} \\ \nonumber %
 &=& {\cal C}_{M,N} {\cal N}_{\vec T}
\,\Delta(\blambda)^2 \prod_{i=1}^N \lambda_i^{M-N} \sum_{\vec m}
\frac{\alpha_{\vec m}(-1)}{d_{\vec m}} \chi_{\vec m}\left({\bf
T}^{-1}\right)
 \chi_{\vec m}\left(diag(\tilde{\blambda}) \right)
\end{eqnarray}

We may now use Weyl's character formula for the unitary groups
(\ref{eq:Weyl_formula_def}) to represent the characters in terms
of the eigenvalues of the corresponding matrices. We also use
(\ref{eq:alpha_x_definition}) to represent $\alpha_{\vec
m}/d_{\vec m}$. As a result of these substitutions we get
\begin{eqnarray} \label{eq:P_lambda_i_semicorr_applied_weyl}
P(  \{\lambda_i\}) = \tilde{C}_{M,N} {\cal N}_{\vec T}
\,\Delta(\blambda)^2 \prod_{i=1}^N \lambda_i^{M-N} \sum_{\vec m}
\left[\prod_{i=1}^\nt\frac{(-1)}{(m_i-i+\nt)!}\right]
\frac{\det\left(t_i^{m_j-j+\nt}\right) \,
\det\left(\tilde{\lambda}_i^{m_j-j+\nt}\right)}{\Delta(\vec
t)\Delta(\tilde{\blambda})}
\end{eqnarray}
where the vector $\vec t$ was introduced with
\begin{equation}\label{eq:t_vec_def}
\vec t = diag\left(eigs\left[\vec T^{-1}\right]\right)
\end{equation}
where $eigs[\vec T^{-1}]$ is the vector of eigenvalues of $\vec
T^{-1}$. In addition,
\begin{equation}\label{eq:C_MN_tilde}
  \tilde{C}_{M,N} = {\cal C}_{M,N}\prod_{i=1}^\nt (i-1)!
\end{equation}
{}From the form of (\ref{eq:P_lambda_i_semicorr_applied_weyl}) it
is clear that the change of summation variables from $m_i$ to
$k_i$ with
\begin{equation}\label{eq:k_eq_m-i+M_def}
  k_i = m_i -i +\nt
\end{equation}
will simplify the notation. Thus the summation over the new
variables is over all integer vectors $\vec k=[k_1, \ldots,
k_\nt]$, with $k_1>k_2>\ldots>k_\nt\geq 0$. The resulting form of
the joint probability distribution $P(  \{\lambda_i\})$ is
\begin{equation} \label{eq:P_lambda_i_semicorr_applied_k_substitution}
P(  \{\lambda_i\}) = \tilde{C}_{M,N} {\cal N}_{\vec T}
\,\Delta(\blambda)^2 \prod_{i=1}^N\lambda_i^{M-N}
 (-1)^{\nt(\nt-1)/2}\sum_{\vec k} \left[\prod_{i=1}^\nt
\frac{\left(-1\right)^{k_i}}{k_i!}\right] \frac{\det(t_i^{k_j}) \,
\det(\tilde{\lambda}_i^{k_j})}{ \Delta(\vec
t)\Delta(\tilde{\blambda})}
\end{equation}

{\bf Step 2: Resummation of series}. The second step in
calculating $P(\{\lambda_i\})$ is to re-sum the series in
(\ref{eq:P_lambda_i_semicorr_applied_k_substitution}). Despite the
apparent complexity of this multiple sum, it is in precisely of
the form of the Cauchy-Binet formula, described in Lemma
\ref{lemma:Cauchy-Binet Formula}. Applying this formula with
$w(k)=(-1)^k/k!$ and therefore $W(x) = \exp(-x)$ to
(\ref{eq:P_lambda_i_semicorr_applied_k_substitution}) we obtain
the joint eigenvalue density of eigenvalues of $\bG\bG^\dagger$
for the case of semicorrelated zero-mean channels:
\begin{equation} \label{eq:P_lambda_i_semicorr_MN_gen}
P(  \{\lambda_i\}) = \tilde{C}_{M,N} \prod_{i=1}^{\nt} t_i^\nr
\,\Delta(\blambda)^2 \prod_{i=1}^N\lambda_i^{M-N}
 (-1)^{\nt(\nt-1)/2}
\frac{\det\left[e^{-t_i\tilde{\lambda}_j}\right]}{\Delta(\tilde{\blambda})\Delta(\vec
t)}
\end{equation}

{\bf Step 3: Allowing for $M>N$}. The final step in the
calculation is to reassign $\tilde{\lambda}_j=\lambda_j$ for
$j=1,\ldots,N$ and $\tilde{\lambda}_j=0$ for $j=N+1,\ldots,M$.

\subsubsection{Case $\nt=N\leq M=\nr$}

In this case (\ref{eq:P_lambda_i_semicorr_MN_gen}) is essentially
the final result with $\tilde{\lambda}_j=\lambda_j$, for
$j=1,\ldots,N$. After some cancellations we have
\begin{equation} \label{eq:P_lambda_i_semicorr_nt_less_nr}
P(  \{\lambda_i\}) = \frac{1}{\nt!}\prod_{j=1}^{\nt} \frac{t_j^\nr
\lambda_j^{\nr-\nt}}{(\nr-\nt+j-1)!} \,\Delta(\blambda)
 (-1)^{\nt(\nt-1)/2}
\frac{\det\left[e^{- t_i\lambda_j}\right]}{\Delta(\vec t)}
\end{equation}

\subsubsection{Case $\nt=M>N=\nr$}

To take the limit $\tilde{\lambda}_{N+1},\ldots,\tilde{\lambda}_M
\rightarrow 0$ we use Lemma \ref{lemma:vandermonde_finite_x0},
which is an application of the l'Hospital rule and appears in
Appendix \ref{app:Proof of Lemma Vandermonde_finite_x0}. Applying
this lemma to (\ref{eq:P_lambda_i_semicorr_MN_gen}) with $f_i(x) =
e^{-t_i x}$  we get the joint probability density
$P(\{\lambda_i\})$ for $\nt>\nr$
\begin{eqnarray} \label{eq:P_lambda_i_semicorr_nt_greater_nr}
P(  \{\lambda_i\}) = \prod_{j=1}^{\nr} \frac{1}{j!}
\prod_{i=1}^{\nt} t_i^\nr \,\Delta(\blambda)
 (-1)^{\nr(\nr-1)/2}
\frac{\det\left[1 \,;\, t_i
\,;\,\ldots\,;t_i^{\nt-\nr-1};e^{-t_i\lambda_1};\ldots;e^{-t_i\lambda_\nr}
\right]}{\Delta(\vec t)}
\end{eqnarray}

Clearly, if we interchange the correlated side from transmitter to
receiver, one only needs to interchange $\nr$ with $\nt$ in the
above equations, where again $t_i$ are the eigenvalues of $\vec
T^{-1}$, see (\ref{eq:t_vec_def}).

Summarizing the results of this section, we have calculated the
joint eigenvalue distribution for $\bG\bG^\dagger$ with $\bG$
being semicorrelated, using a single unified approach, rather than
two independent methods used before, for the two limits $\nt>\nr$
\cite{Gao2000_DeterminantRepresentationForDistributionOfQuadraticForms}
and $\nr>\nt$
\cite{Chiani2003_CapacityOfSpatiallyCorrelatedMIMOChannels}.\footnote{We
note that had
\cite{Chiani2003_CapacityOfSpatiallyCorrelatedMIMOChannels}
applied Lemma \ref{lemma:vandermonde_finite_x0} to their result,
they would also had been able to get both $\nt>\nr$ and $\nt<\nr$
results.} In addition, our method, based on the character
expansion (Lemma \ref{lemma:Character_Expansion}), the
Cauchy-Binet formula (Lemma \ref{lemma:Cauchy-Binet Formula}), and
the generalized l'Hospital rule (Lemma
\ref{lemma:vandermonde_finite_x0}) is simple enough to be applied
to other types of problems, as we shall see below.

To obtain the moment generating function $g(z)$ one needs to
integrate over the $\lambda$'s, see (\ref{eq:g_z_lambda}). From
this point one can employ the methods by
\cite{Chiani2003_CapacityOfSpatiallyCorrelatedMIMOChannels,
Smith2003_CapacityMIMOsystems_with_semicorrelated_flat_fading} to
calculate the moment generating function. For completeness, we
include the calculation of  $g(z)$ in Appendix \ref{app:Moment
generating function for semicorrelated and non-zero mean channels}
for both cases above.

\subsection{Case of uncorrelated $\bG$ with non-zero mean $\bG_0$}
\label{sec:Case of uncorrelated G with non-zero mean G0}

We will now analyze a simple case, where both unitary integrals
are present in (\ref{eq:P_lambda_i_int1}), namely the case of
uncorrelated $\bG$ with non-zero mean $\bG_0$. In this case the
probability density $p(\bG)$ takes the form
\begin{eqnarray}\label{eq:p_G_non_zero_mean_def}
  p(\bG) &=&
  \exp\left[-\Tr{\left(\bG-\bG_0\right)^\dagger\left(\bG-\bG_0\right)}\right]
  \\ \label{eq:p_G_non_zero_mean_def2}
  &=& e^{-\sum_{i=1}^N\lambda_i-\sum_{i=1}^{N_0}
  \gamma_i} \exp\left[\Tr{\bG_0^\dagger \bU  diag(\bmu) \bV^\dagger} + \Tr{ \bV  diag(\bmu)^\dagger\bU^\dagger\bG_0}\right]
\end{eqnarray}
where $\gamma_i$ for $i=1,\ldots,N_0$ are the non-zero eigenvalues
of $\bG_0^\dagger\bG_0$ and where we have used the singular value
decomposition of $\bG$, (\ref{eq:svd_def}).

{\bf Step 1: Integration over $\vec U$, $\vec V$}. It will be
convenient below to have the integrations of both $\bU$ and $\bV$
over the same unitary group, therefore we will extend both $\bG_0$
and $\bG$ in the second line in (\ref{eq:p_G_non_zero_mean_def2})
to $M\times M$ matrices, by adding an appropriate number of
columns or rows, which will eventually be set to zero. Essentially
what we are doing is allowing initially for $M$ non-zero
eigenvalues of $\bG\bG^\dagger$ for the terms in the exponent and
later, if appropriate ($M\neq N$), letting the extra ones go to
zero. Inserting (\ref{eq:p_G_non_zero_mean_def}) into
(\ref{eq:P_lambda_i_int1}) we get
\begin{eqnarray}\label{eq:P_lambda_non_zero_mean_int}
    P(  \{\lambda_i\}) &=& {\cal C}_{M,N} \Delta(\blambda)^2 \, \prod_{i=1}^{N_0}e^{-\gamma_i}\, \prod_{i=1}^N
   \left[\lambda_i^{M-N} e^{-\lambda_i}\right] \\ \nonumber
  &\cdot& \int D\bU \int  D\bV \,\,
  \exp\left[\Tr{{\bG_0}^\dagger { \bU  diag(\bmu)  \bV}^\dagger} + \Tr{\bV
   diag(\bmu)^\dagger\bm\bU^\dagger\bm\bG_0}\right]
\end{eqnarray}
We may now apply the character expansion Lemma
\ref{lemma:Character_Expansion} to each of the exponentials of
traces above. As a result, the second line in the equation above
becomes
\begin{equation}\label{eq:P_lambda_nzero_mean_double_int}
   \int D\bU \int D\bV \, \sum_{\vec m}\sum_{\vec m'}
    \alpha_{\vec m}(1) \alpha_{\vec m'}(1)
    \chi_{\vec m}\left( \bG_0^\dagger \bU  diag(\bmu) \bV^\dagger \right)
    \chi_{\vec m'}\left( \bV  diag(\bmu)^\dagger \bU^\dagger
    \bG_0\right)
\end{equation}
Since both $\bV$ and $\bU$ appear as quadratic forms in the double
sum, we may apply (\ref{eq:def_integration_over_U_U_bar}) to
integrate both unitary integrals.
(\ref{eq:P_lambda_nzero_mean_double_int}) thus becomes
\begin{equation}\label{eq:P_lambda_nzero_mean_int_UV}
   \sum_{\vec m}
    \frac{\alpha_{\vec m}(1)^2}{d_{\vec m}^2}
    \chi_{\vec m}\left( \bG_0^\dagger \bG_0\right) \chi_{\vec m}
    \left(  diag(\tilde{\blambda})\right)
    = \sum_{\vec k} \left[\frac{\prod_{j=1}^M (j-1)!}{\prod_{j=1}^M
    k_j!}\right]^2 \left[\frac{\det(\tilde{\gamma}_i^{k_j}) \,
\det(\tilde{\lambda}_i^{k_j})}{
\Delta(\tilde{\bgamma})\Delta(\tilde{\blambda})}\right]
\end{equation}
where in the right hand side of the above equation we applied
Weyl's character formula (\ref{eq:Weyl_formula_def}), substituted
for $\alpha_{\vec m}$ using (\ref{eq:alpha_x_definition}) and
redefined the summation in terms of $\vec k$
(\ref{eq:k_eq_m-i+M_def}). Once again we introduced the
$M$-dimensional vector $\tilde{\blambda}$ with elements
$\tilde{\lambda}_i=\lambda_i$ for $i=1,\ldots,N$ and (for $M>N$)
$\tilde{\lambda}_{N+1},\ldots,\tilde{\lambda}_M$ arbitrary at this
point. In addition, we introduced another $M$-dimensional vector,
$\tilde{\bgamma}$, such that $\tilde{\gamma}_i=\gamma_i$ for
$i=1,\ldots,N_0$ and, (assuming $N_0<M$) for $i=N_0+1,\ldots,M$ it
is currently arbitrary, but will eventually be set to zero.

{\bf Step 2: Resummation of series}. We next apply the
Cauchy-Binet formula, Lemma \ref{lemma:Cauchy-Binet Formula} to
resum the above sums, using $w(k)=1/k!^2$ and the Taylor expansion
of the modified Bessel function \cite{Abramowitz_Stegun_book}
\begin{equation}\label{eq:Bessel_modified_expansion}
  I_0(x) = \sum_{k=0}^\infty \frac{1}{(k!)^2}
  \left[\frac{x}{2}\right]^{2k}
\end{equation}
(\ref{eq:P_lambda_nzero_mean_int_UV}) thus becomes
\begin{equation}\label{eq:nzero_mean_applied_Cauchy_Binet}
  \prod_{j=1}^M
  (j-1)!^2\frac{\det\left(I_0(2\sqrt{\tilde{\gamma}_i\tilde{\lambda}_j})\right)}{\Delta(\tilde{\bgamma})
  \Delta(\tilde{\blambda})}
\end{equation}
Inserting this into (\ref{eq:P_lambda_non_zero_mean_int}) we
obtain
\begin{eqnarray}\label{eq:P_lambda_non_zero_mean_fin1}
    P(  \{\lambda_i\}) = \frac{\prod_{j=1}^{M-N}\left[(j-1)!(N+j-1)!\right]}{N!} \Delta(\blambda)^2 \,
    \prod_{i=1}^{N_0}e^{-\gamma_i}\, \prod_{i=1}^N
   \left[\lambda_i^{M-N} e^{-\lambda_i}\right] 
  \frac{\det\left(I_0(2\sqrt{\tilde{\gamma}_i\tilde{\lambda}_j})\right)}{\Delta(\tilde{\bgamma})
  \Delta(\tilde{\blambda})}
\end{eqnarray}

{\bf Step 3: Allowing for $M> N$ and $M> N_0$}. As in the previous
section, we may now let $\tilde{\gamma}_i=\gamma_i$ for
$i=1,\ldots,N_0$ and $\tilde{\lambda}_i=\lambda_i$ for
$i=1,\ldots,N$ and if $M>N$ and/or $M>N_0$ we may apply Lemma
\ref{lemma:vandermonde_finite_x0} with the parameter $x_0=0$ to
let the superfluous $\tilde{\gamma}_i$ for $i=N_0+1,\ldots,M$ and
$\tilde{\lambda}_i$ for $i=N+1,\ldots,M$ go to zero. The final
expression for $P(\{\lambda_i\})$ reads
\begin{eqnarray}\label{eq:P_lambda_non_zero_mean_fin2}
    P(  \{\lambda_i\}) =  \frac{(-1)^{(N_0+N)(M-1)}}{N!}\frac{\Delta(\blambda) \det\vec L}{\Delta(\bgamma)}  \,\, %
\prod_{i=1}^Ne^{-\lambda_i}\prod_{j=1}^{N_0} e^{-\gamma_j}
\frac{%
\prod_{i=1}^{M-N}\left[\frac{(N+i-1)!}{(i-1)!}\right] \,
} { %
\prod_{j=1}^{M-N_0}\left[(j-1)!\right]^2 \, \prod_{j=1}^{N_0}
\gamma_j^{M-N_0} %
}
\end{eqnarray}
where it is assumed that in the last ratio the numerator is unity
when $M=N$, while the denominator is unity when $M=N_0$. $\vec L$
is a $M\times M$ matrix with elements
\begin{equation}\label{eq:nzmean_Lmatrix}
  L_{ij} = \left\{
\begin{array}{cc}
  I_0(2\sqrt{\gamma_i\lambda_j}) & i\leq N_0, j\leq N   \\
  \lambda_j^{i-N_0-1} & i> N_0, j\leq N  \\
  \gamma_i^{j-N-1} & i\leq N_0, j>N  \\
  \left((i-N_0-1)!\right)^2\delta_{i-N_0,j-N} & i> N_0, j> N
\end{array}
\right.
\end{equation}
This obtains the result by
\cite{Kang2002_CapacityOfMIMORicianChannels} using simple Bessel
functions rather than more complicated hypergeometric functions,
with only a few simple steps involved. Using this result, we
calculate $g(z)$ in Appendix \ref{app:Moment generating function
for semicorrelated and non-zero mean channels}.

\subsection{Case of fully correlated $\bG$ with zero mean}
\label{sec:Case of fully correlated G with zero mean}

In the previous sections we recovered already known results,
albeit in a very direct and simple way.  We now turn to a problem
that has not been previously solved.   In this section we will
analyze the joint probability distribution of eigenvalues of
$\bG^\dagger\bG$ when $\bG$ has zero mean but is fully correlated,
i.e. its probability density is
\begin{equation}\label{eq:p_G_fully_corr_zero_mean}
  p(\bG) = {\cal N}_{\vec T,\vec R} \exp\left[-\Tr{\bG^\dagger
  \vec T^{-1} \bG \vec R^{-1}}\right]
\end{equation}
with
\begin{equation}\label{eq:N_TR_def}
  {\cal N}_{\vec T, \vec R}^{-1} = \det\vec T^\nr \det\vec R^\nt
\end{equation}
Applying (\ref{eq:GeqUmuV_def}) to
(\ref{eq:p_G_fully_corr_zero_mean}) and then inserting it  into
(\ref{eq:P_lambda_i_int1}) we get
\begin{eqnarray}\label{eq:P_lambda_i_fully_corr}
    P(  \{\lambda_i\}) &=& {\cal N}_{\vec T,\vec R} \, {\cal C}_{M,N} \Delta(\blambda)^2 \!\int\! D\bU \!\int\! D\bV
\exp\left[-\Tr{ {\vec T^{-1} \bU  diag(\bmu) \bV^\dagger \vec
R^{-1} \bV  diag(\bmu)^\dagger
\bU^\dagger}}\right] \nonumber \\ %
&=& {\cal N}_{\vec T,\vec R} \, \Delta(\blambda) {\cal I}(\vec T,
\vec R,\blambda)
\end{eqnarray}
where the second equation defines the quantity ${\cal I}(\vec
T,\vec R,\blambda)$. At this point it is convenient to assume that
$M=N$, relaxing the constraint later.

{\bf Step 1: Integration over $\bU$, $\bV$}. We may now use Lemma
\ref{lemma:Character_Expansion} to expand the exponential so that
the integration over $\bU$, $\bV$ can be performed. As a result we
get
\begin{eqnarray} \label{eq:P_lambda_i_fullycorr_char_exp_int_over_UV}
{\cal I}(\vec T, \vec R,\blambda) &=& {\cal C}_{M,N}
\Delta(\blambda) \sum_{\vec m} \frac{\alpha_{\vec m}(-1)}{d_{\vec
m}^2} \chi_{\vec m}\left({\bf T}^{-1}\right)
 \chi_{\vec m}\left(\blambda \right) \chi_{\vec m}\left({\bf R}^{-1}\right)
\end{eqnarray}
where $\alpha_{\vec m}$ is the expansion coefficient, see
(\ref{eq:alpha_x_definition}), and $d_{\vec m}$ is the dimension
of the $\vec m$ representation, see (\ref{eq:d_r_definition}).
Comparing the above equation to
(\ref{eq:P_lambda_i_semicorr_char_exp_int_over_U}),
(\ref{eq:P_lambda_nzero_mean_int_UV}) we see a crucial difference
here, namely that $\alpha_{\vec m}$, $d_{\vec m}$ do not appear
with the  same powers. This is important because, using
(\ref{eq:alpha_x_definition}) we can no longer cancel $d_{\vec
m}$, which as seen (\ref{eq:d_r_definition}) is a rather
complicated function. Applying Weyl's character formula
(\ref{eq:Weyl_formula_def}) and redefining the summations from
$\vec m$ to $\vec k$, where $k_i = m_i -i +M$ similar to
(\ref{eq:k_eq_m-i+M_def}) we get
\begin{equation} \label{eq:P_lambda_i_fullycorr_Applied_k_substitution}
{\cal I}(\vec T, \vec R,\blambda) = \frac{1}{M!}\sum_{\vec k}
\left[\prod_{i=1}^M \frac{\left(-1\right)^{k_i}}{k_i!}\right]
 \frac{\det(t_i^{k_j})
\, \det(\lambda_i^{k_j}) \, \det(r_i^{k_j})}{\Delta(\vec
k)\Delta(\vec t)\Delta(\vec r)}
\end{equation}
where we also used Lemma \ref{lemma:Vandermonde determinant form
of d_r} to express $d_{\vec k}$ in terms of $\Delta(\vec k)$ and
we introduced the vectors $\vec t$, $\vec r$, with elements the
inverse eigenvalues of $\vec T$, $\vec R$, respectively, i.e.
\begin{eqnarray}\label{eq:t_vec_r_vec_def}
  \vec t &=& diag\left(eigs\left[\vec T^{-1}\right]\right) \\
  \nonumber
\vec r &=& diag\left(eigs\left[\vec R^{-1}\right]\right)
\end{eqnarray}

Unfortunately, we may not make any further progress in resumming
the above expression, due to two impediments. First, to apply
Cauchy-Binet's formula, Lemma \ref{lemma:Cauchy-Binet Formula}, we
need two determinants of the form $\det(x_i^{k_j})$ in the
numerator. Instead we have three. In addition, the extra
$\Delta(\vec k)$ in the denominator does not allow us to separate
the different $k_i$ sums. However, we can still get the moment
generating function for the mutual information $g(z)$, by first
integrating over $\lambda_i$. This will be done in the next
section.

\section{Moment-Generating Function for $\bG$ fully correlated}
\label{sec:Moment-Generating Function with no Interference}

Even if in Section \ref{sec:Case of fully correlated G with zero
mean} we were not able to find an final expression for the joint
probability of eigenvalues of $\bG^\dagger\bG$ for the case of
fully correlated $\bG$, in this section we will generate a
closed-form expression for the moment generating function of the
mutual information, $g(z)$, using
(\ref{eq:P_lambda_i_fullycorr_Applied_k_substitution}).

\subsection{$g(z)$ for the case $M=N$}
\label{sec:g_z_for M_eq_N}

 From (\ref{eq:g_z_lambda}) $g(z)$ can
be written (for $M=N$) as
\begin{equation}\label{eq:g_z_diagonal_def}
g(z) = \int d\blambda   \prod_{i=1}^M    (1 + \lambda_i)^z
P(\{\lambda_i\})
\end{equation}
where the integral is over positive $\lambda_i$'s, and $P$ is the
joint probability distribution function of the eigenvalues.
Inserting (\ref{eq:P_lambda_i_fullycorr_Applied_k_substitution})
into (\ref{eq:P_lambda_i_fully_corr}) $g(z)$ can be written as
\begin{eqnarray} \label{eq:g_z_with_sums1}
g(z) &=& \frac{{\cal N}_{\vec T,\vec R}}{M!\Delta(\vec t)
\Delta(\vec r)} \int d {\blambda} \prod_{i=1}^M (1 + \lambda_i)^z
 \Delta(\blambda) \sum_{\vec k} \left[\prod_{i=1}^M \frac{\left(-1\right)^{k_i}}{k_i!}\right] \frac{\det(t_i^{k_j}) \det(r_i^{k_j})
    \det(\lambda_i^{k_j})}{\Delta(\vec k)}
\end{eqnarray}

{\bf Step 2: Integration over $\{\lambda_i\}$}. As mentioned
above, we cannot resum the series in the form
(\ref{eq:g_z_with_sums1}). Instead, we will first integrate over
$\lambda_i$. To proceed, it is useful to transform the Vandermonde
determinant $\Delta(\blambda)$ into another one with arguments
$\lambda_i/(1+\lambda_i)$
\begin{eqnarray}\label{eq:Delta(lambda_over_1+lambda}
    \Delta(\blambda) &=& \prod_{i>j}[\lambda_i-\lambda_j] =
    \prod_{i>j}\left[(1+\lambda_i)-(1+\lambda_j)\right] \\
    \nonumber
    &=& \left[\prod_{m=1}^M (1+\lambda_m)^{M-1} \right]\prod_{i>j}\left(\frac{1}{1+\lambda_j}-\frac{1}{1+\lambda_i}\right) \\
    \nonumber
    &=& \left[\prod_{m=1}^M (1+\lambda_m)^{M-1}\right] \prod_{i>j}\left(\frac{\lambda_i}{1+\lambda_i}-\frac{\lambda_j}{1+\lambda_j}\right) \\
    \nonumber
    &=& \left[\prod_{m=1}^M (1+\lambda_m)^{M-1}\right]
    \det\left[\left(\frac{\lambda_i}{1+\lambda_i}\right)^{j-1}\right]
\end{eqnarray}
The usefulness of this form of the determinant will become
apparent in a moment. We first expand the above determinant using
$\sgn(\vec b)$ as in (\ref{eq:det_expansion_def}). Thus $g(z)$
becomes
\begin{equation}\label{eq:g_z_with_epsilon_b_i}
  g(z) = {\cal N}_{{\bf T R}} \sum_{\vec b} \sgn(\vec b) \int
  d\blambda \left\{\prod_{i=1}^M (1+\lambda_i)^{z+M-b_i}\right\}
  \left[ \prod_{j=1}^M \lambda_j^{b_j-1} {\cal I}(\vec T,  \vec R,
  \blambda)\right]
\end{equation}
We now wish to integrate the above expression by parts with
respect to the $\lambda$'s, integrating the curly brackets and
differentiating the square brackets.  Usually, when integrating by
parts we generate various boundary terms evaluated at $\lambda=0$
and $\lambda=\infty$. For example, if we integrate by parts $p$
times over a particular $\lambda$, we will get a sum over $p$
terms, in which the $n$th term (with $n=1,\ldots,p$) has the
square brackets differentiated $n-1$ times with respect to this
$\lambda$ and is evaluated at $\lambda=0$ and $\lambda=\infty$.
However, here we will point out that these boundary terms are zero
in the cases we are concerned with.

We note that according to Lemma \ref{lemma:I_exponentially
convergent} in Appendix \ref{app:Proof of Lemma: I_exponentially
convergent} the quantity ${\cal I}(\vec T, \vec R, \blambda)$ is
exponentially convergent when any $\lambda_i$ approaches infinity.
Therefore, the square bracketed quantity above, as well as any of
its derivatives with respect to the $\lambda$'s is zero when
evaluated at any $\lambda_i=\infty$. As a result, we may discard
all generated boundary terms at $\lambda_i=\infty$. In addition,
we observe from its definition in
(\ref{eq:P_lambda_i_fullycorr_Applied_k_substitution}) that ${\cal
I}$ is finite when any of the $\lambda$'s approach zero. Thus if
we integrate each $\lambda_i$ up to $b_i-1$ times by parts in
(\ref{eq:g_z_with_epsilon_b_i}), all boundary terms evaluated at
$\lambda_i=0$ for $i=1,\ldots,M$, will vanish, since they involve
derivatives of the square-bracketed quantity above, not exceeding
$b_i-1$ order.  Indeed, we will choose to integrate by parts
precisely $b_i -1$ times in order to make further progress.

To proceed with this multiple integration by parts, we first note
that the $(b_i-1)$th indefinite integral of the curly brackets in
(\ref{eq:g_z_with_epsilon_b_i}) with respect to all $\lambda_i$ is
\begin{equation}\label{eq:b_i-1_indef_int_curly}
  \prod_{i=1}^M \frac{(1+\lambda_i)^{z+M-1}}{\prod_{p=1}^{b_i-1}
  (z+M-p)}
\end{equation}
with the product in the denominator assumed to be unity for
$b_i=1$. To obtain the $(b_i-1)$th derivative of the square
brackets in (\ref{eq:g_z_with_epsilon_b_i}) we need to rewrite
${\cal I}$ as an expansion in the form of
(\ref{eq:P_lambda_i_fullycorr_Applied_k_substitution}). In
particular, using $\sgn(\vec c)$ (see
(\ref{eq:det_expansion_def})) to expand $\det(\lambda_i^{k_j})$ in
(\ref{eq:P_lambda_i_fullycorr_Applied_k_substitution}), we can
rewrite the square brackets of (\ref{eq:g_z_with_epsilon_b_i}) as
\begin{equation}\label{eq:square_bracks_eps_expansion}
  \left[\,\cdot\,\right] = \frac{1}{M!}\sum_{\vec k}
\left[\prod_{i=1}^M \frac{\left(-1\right)^{k_i}}{k_i!}\right]
 \frac{\det(t_i^{k_j}) \, \det(r_i^{k_j})}{\Delta(\vec
k)\Delta(\vec t)\Delta(\vec r)} \, %
 \sum_{\vec c} \sgn(\vec c) \prod_{j=1}^M \lambda_j^{k_{c_j} + b_j -1}
\end{equation}
where the sum over $\vec c$ is over all permutations of
$(1,2,\ldots,M)$. By differentiating all $\lambda_i$ in the above
expression $b_i-1$ times we get
\begin{equation}\label{eq:square_bracks_eps_expansion2}
\frac{1}{M!}\sum_{\vec k} \left[\prod_{i=1}^M
\frac{\left(-1\right)^{k_i}}{k_i!}\right]
 \frac{\det(t_i^{k_j}) \, \det(r_i^{k_j})}{\Delta(\vec
k)\Delta(\vec t)\Delta(\vec r)} \, %
 \sum_{\vec c} \sgn(\vec c) \prod_{j=1}^M \left[\lambda_j^{k_{c_j}} \prod_{p=1}^{b_j-1}
(k_{c_j}+p)\right]
\end{equation}
Combining all above results, $g(z)$ of
(\ref{eq:g_z_with_epsilon_b_i}) can be expressed as
\begin{eqnarray}\label{eq:g_z_with_epsilon_b_i_c_i}
  g(z) = \frac{{\cal N}_{{\bf T R}}}{M!}  \int
  d\blambda \prod_{i=1}^M (1+\lambda_i)^{z+M-1}
\sum_{\vec k} && \left[\prod_{i=1}^M
\frac{\left(-1\right)^{k_i}}{k_i!}\right]
 \frac{\det(t_i^{k_j}) \, \det(r_i^{k_j})}{\Delta(\vec
k)\Delta(\vec t)\Delta(\vec r)} \\ \nonumber %
 \times&&\sum_{\vec b,\vec c} \sgn(\vec b) \sgn(\vec c) %
 \prod_{j=1}^M \left[(-1)^{b_j-1}\lambda_j^{k_{c_j}} \prod_{p=1}^{b_j-1}
\frac{k_{c_j}+p}{z+M-p}\right]
\end{eqnarray}
where the additional $(-1)^{b_j-1}$ comes from the integration by
parts.

We will now integrate over the $\lambda_i$'s {\em before} summing
over the $\vec k$'s. Looking at the above equation this seems
problematic since each of the terms in the sum over $\vec k$ above
is unbounded at large $\lambda$. However, from Lemma
\ref{lemma:I_exponentially convergent} in Appendix \ref{app:Proof
of Lemma: I_exponentially convergent} we know that ${\cal I}$ and
thus the integrand above is exponentially bounded at large
$\lambda$ and hence integrable. To circumvent the discrepancy we
introduce a cutoff function $f(\lambda)$ for every $\lambda$,
which is unity at $\lambda\rightarrow 0$ and tends to zero faster
than a power law as $\lambda\rightarrow\infty$. For example, we
can take $f(x)=e^{-\delta x}$. In the end of the calculation we
will set $\delta=0$ and thus make $f(x)$ unity for all $x$.
Cutting off the integrals makes all terms finite and thus we can
freely interchange the order of the summation and integration.
Thus we have
\begin{eqnarray}\label{eq:g_z_with_epsilon_b_i_c_i_interchange_int_sum}
  g(z) &=& \frac{{\cal N}_{{\bf T R}}}{M!}
\sum_{\vec k} \left[\prod_{i=1}^M
\frac{\left(-1\right)^{k_i}}{k_i!}\right]
 \frac{\det(t_i^{k_j}) \, \det(r_i^{k_j})}{\Delta(\vec
k)\Delta(\vec t)\Delta(\vec r)} \\ \nonumber %
 &&\times\sum_{\vec b,\vec c} \sgn(\vec b) \sgn(\vec c) %
 \prod_{i=1}^M \left\{(-1)^{b_i-1}\prod_{p=1}^{b_i-1}
\frac{k_{c_i}+p}{z+M-p} \int_0^\infty d\lambda_i
\lambda_i^{k_{c_i}} (1+\lambda_i)^{z+M-1} f(\lambda_i)\right\}
\end{eqnarray}
Next it is important to appreciate the following property of the
second line of the above equation. Once the integrals over
$\lambda_i$ have been performed,  the quantities in the curly
bracket depend on the index $i$ only through the indices $b_i$ and
$c_i$. Therefore, the second line of the above equation is in the
form of (\ref{eq:det_expansion_def}) and thus it can be written as
$M!$ times a determinant of the matrix $\vec W$ with elements
\begin{eqnarray}\label{eq:W_ij_summed_over_eb_ec}
  W_{ij} &=& \prod_{p=1}^{i-1} \frac{k_j+p}{z+M-p} (-1)^{i-1}
  \int_0^\infty d\lambda \lambda^{k_j} (1+\lambda)^{z+M-1}
  f(\lambda) \\ \nonumber
  &=& \prod_{p=1}^{i-1} \frac{k_j+p}{z+M-p} (-1)^{i-1} W_{1j}
\end{eqnarray}
where the second equation results from the fact that the only
$i$-dependence of $W_{ij}$ is outside the integral. Thus $W_{1j}$
is simply the integral above and can be factored outside the
determinant, resulting in
\begin{eqnarray}\label{eq:detW_ij_summed_over_eb_ec}
  \det \vec W &=& \prod_{j=1}^M W_{1j} \det\left[\prod_{p=1}^{i-1} \frac{k_j+p}{z+M-p}
  (-1)^{i-1}\right] \\ \label{eq:detW_ij_summed_over_eb_ec2}
  &=& (-1)^{M(M-1)/2} \prod_{j=1}^M \frac{W_{1j}}{(z+j-1)^{j-1}}
  \det\left(\frac{(k_j+i-1)!}{k_j!}\right)
\end{eqnarray}
In (\ref{eq:detW_ij_summed_over_eb_ec2}), the power of $(-1)$
comes from bringing $(-1)^{i-1}$ outside the determinant, while
the product of $(z+j-1)^{j-1}$ originates from the denominator in
the determinant.

Using the same procedure as in Lemma \ref{lemma:Vandermonde
determinant form of d_r} in Appendix \ref{app:Proof of Lemma
Vandermonde determinant form of d_r}, we can show that the
determinant as in the right-hand side of
(\ref{eq:detW_ij_summed_over_eb_ec2}) is simply equal to
$(-1)^{M(M-1)/2}\Delta(\vec k)$. We can thus rewrite the second
line in (\ref{eq:g_z_with_epsilon_b_i_c_i_interchange_int_sum}) as
\begin{equation}\label{eq:second_term_in_g_z_interchanged_int_sum}
  \prod_{i=1}^M \frac{\int_0^\infty d\lambda \lambda^{k_i}
  (1+\lambda)^{z+M-1} f(\lambda)}{(z+i-1)^{i-1}} \Delta(\vec k)
\end{equation}
Inserting this into
(\ref{eq:g_z_with_epsilon_b_i_c_i_interchange_int_sum}) we see
immediately that $\Delta(\vec k)$ cancels leaving
\begin{eqnarray}\label{eq:g_z_with_epsilon_b_i_c_i_interchange_int_sum2}
  g(z) = {\cal N}_{{\bf T R}} \prod_{i=1}^M \frac{1}{(z+i-1)^{i-1}}
\sum_{\vec k}  \frac{\det(t_i^{k_j}) \,
\det(r_i^{k_j})}{\Delta(\vec t)\Delta(\vec
  r)} %
\prod_{i=1}^M \left[\int_0^\infty d\lambda
\frac{(-\lambda)^{k_i}}{k_j!}
  (1+\lambda)^{z+M-1} f(\lambda) \right]
\end{eqnarray}

{\bf Step 3: Resummation of series}. We may now use the
Cauchy-Binet theorem (see Lemma \ref{lemma:Cauchy-Binet Formula})
to the above form of the sum over $\vec k$, which gives
\begin{eqnarray}\label{eq:sum_k_final_form_cauchy}
  &&\det\left[\int_0^\infty d\lambda e^{-r_it_j\lambda}
  (1+\lambda)^{z+M-1} f(\lambda)\right] \\ \nonumber
  &=& \det\left[\int_0^\infty d\lambda e^{-r_it_j\lambda}
  (1+\lambda)^{z+M-1}\right] \\
  &=& \prod_{i=1}^M \frac{1}{(t_i r_i)^M} \det F(r_it_j;z)
\label{eq:sum_k_final_form_cauchy3}
\end{eqnarray}
In the second line above we discarded the function $f(x)$, since
the integral is now convergent.  I.e., we now take the limit $f(x)
= \lim_{\delta \rightarrow 0} e^{-\delta x} \rightarrow 1$. Thus,
$F(x,z)$ is given by \cite{Gradshteyn_Ryzhik_book}
\begin{equation}\label{eq:F_x_z}
  F(x,z) = x^M \int_0^\infty d\lambda e^{-x\lambda}
  (1+\lambda)^{z+M-1} = x^{-z} e^x \Gamma(z+M,x)
\end{equation}
where $\Gamma(\alpha,x)$ is the incomplete $\Gamma$ function. In
summary, the moment generating function of the mutual information
$g(z)$ can be written for $\nt=\nr=M=N$ as
\begin{equation}\label{eq:g_z_final_M=N}
  g(z) = \frac{1}{\prod_{i=1}^M (z+i-1)^{i-1}\Delta(\vec t)\Delta(\vec
  r)} \det F(t_i r_j; z)
\end{equation}
where the normalization ${\cal N}_{\vec T,\vec R}$ has been
cancelled with the term in front of the determinant in
(\ref{eq:sum_k_final_form_cauchy3}). We note that the above
expression is symmetric in $\vec T$ and $\vec R$ for $M=N$.

\subsection{$g(z)$ for the case $M>N$}
\label{sec:g_z_for M_greater_N}

{\bf Step 4: Allowing for $M>N$}. Next, we will obtain the
expression for $M\neq N$, and specifically, without loss of
generality we will assume that $\nt>\nr$, i.e. $M=\nt$, $N=\nr$.
This can be readily obtained by first assuming that $\vec R$ is
$M$-dimensional with the first $N$ eigenvalues the ones
corresponding to the correlations at the receiver and then taking
the remaining $M-N$ eigenvalues to 0. Thus the $M\neq N$ problem
is obtained by letting the $M-N$ eigenvalues of one of the
correlation matrices go to zero. In the particular situation, we
will let $r_j$, for $j=N+1,\ldots,M$ go to infinity.

Even though both numerator and denominator of
(\ref{eq:g_z_final_M=N}) are unbounded in this case, using Lemma
\ref{lemma:vandermonde_infinite_x} in Appendix \ref{app:Proof of
Lemma Vandermonde_finite_x0} we can see that the ratio is
well-defined and finite. Specifically, we see that the function
$f_i(x_j)$ in the Lemma is $F(t_ir_j,z)$. Noting that the
asymptotic expansion of $F(x,z)$ is (see
\cite{Gradshteyn_Ryzhik_book}, Equation 8.357)
\begin{equation}\label{eq:asymptotic_F}
  F(x,z) \approx x^{M-1} \sum_{k=0}^\infty
  \frac{\Gamma(M+z)}{\Gamma(M+z-k)} x^{-k}
\end{equation}
we find that when $r_{N+1},\ldots,r_M\rightarrow \infty$ the ratio
becomes
\begin{equation}\label{eq:ratio_for_large_r}
\frac{\det F(t_ir_j;z)}{\Delta(\vec r_M)} \rightarrow
\prod_{j=1}^{M-N-1} (M+z-j)^{M-N-j} \frac{\det \vec L}{\Delta(\vec
r_N)}
\end{equation}
where we used the notation $\vec r_N$, $\vec r_M$ to explicitly
specify the length of the vectors and $\vec L$ is a
$M$-dimensional matrix with elements
\begin{equation}\label{eq:Lij}
  L_{ij} = \left\{\begin{array}{cc}
  F(t_ir_j;M+z) & j\leq N \\
  t_i^{j-1} & j>N
  \end{array} \right.
\end{equation}

Summarizing all results above, we thus have the general form of
the moment generating function $g(z)$ of the mutual information
$I$ for general $M$, $N$ which is
\begin{equation}\label{eq:g_z_final_MneqN}
  g(z) = E_{\bG}\left[e^{zI}\right] =
  \frac{\prod_{j=1}^{M-N-1} (M+z-j)^{M-N-j}}{\prod_{i=1}^M (z+i-1)^{i-1}}
  \frac{\det\vec L}{\Delta(\vec t)\Delta(\vec r)}
\end{equation}
with the elements of $\vec L$ given by (\ref{eq:Lij}) for
$\nt\geq\nr$ and with $t_i$ and $r_i$ interchanged for $\nr>\nt$.

\section{Ergodic Capacity for fully correlated channels}
\label{sec:Ergodic Capacity for fully correlated channels}

In this section we will derive an expression for the ergodic
capacity in the case of fully correlated channels discussed in the
previous section. Without loss of generality, we will assume that
$\nr=N\leq M=\nt$, as is implicitly assumed in
(\ref{eq:g_z_final_MneqN}). We apply (\ref{eq:g_z_derivative1})
and  differentiate $g(z)$ in (\ref{eq:g_z_final_MneqN}) with
respect to $z$:
\begin{equation}\label{eq:g_z_first_derivative}
  g'(z) = g(z) \left(\sum_{j=1}^{M-N-1} \frac{M-N-j}{M+z-j} -
  \sum_{i=1}^{M-1} \frac{i}{z+i} + \frac{1}{\det\vec L}
  \frac{\partial \det \vec L }{\partial z}\right)
\end{equation}
We next have to set $z=0$, to get $E[I]$, as in
(\ref{eq:g_z_derivative1}), recalling that $g(0)=1$, see
(\ref{eq:g_G_def}). This results to
\begin{eqnarray}\label{eq:ergodic_cap}
E[I] = g'(0) = \sum_{j=1}^{M-N-1} \frac{j}{N+j}  +\sum_{j=1}^N
\frac{\det \tilde{\vec L}_j}{\det\vec L} -M+1
\end{eqnarray}
where the matrix $\tilde{\vec L}_j$ is the matrix $\vec L$ with
its $j$-th column replaced by the column $D_{1,ij}$, i.e.
\begin{equation}\label{eq:L_tilde_def}
  \tilde{\vec L}_j = \left[L_{i,1}; L_{i,2}; \ldots; L_{i,j-1}; D_{1,ij};
  L_{i,j+1}; \ldots\right]
\end{equation}
where the element $D_{i,j}$ is given by
\begin{eqnarray}\label{eq:D_ij_def}
  D_{ij} &=& \left.\frac{\partial F(t_ir_j;M+z)}{\partial z}\right|_{z=0}
  \\ \nonumber
 &=& \left.\left( e^x x^M (-1)^M \left[\frac{Ei(-x)}{x}\right]^{(M-1)}
 \right)\right|_{x=t_ir_j}
\end{eqnarray}
where the superscript $(M-1)$ in the square brackets denotes the
$M-1$st derivative and $Ei(x)$ is the exponential integral
\cite{Gradshteyn_Ryzhik_book}. Note that the second sum in
(\ref{eq:ergodic_cap}) goes up to $N$, since for $j>N$ the columns
in $\vec L$ do not depend on $z$, as seen in
(\ref{eq:g_z_final_MneqN}). Similarly one can calculate higher
moments of the mutual information distribution.

\subsection{Example}
\label{sec:example_ergod_cap}

\begin{figure}[htb]
\centerline{\epsfxsize=.9\columnwidth\epsffile{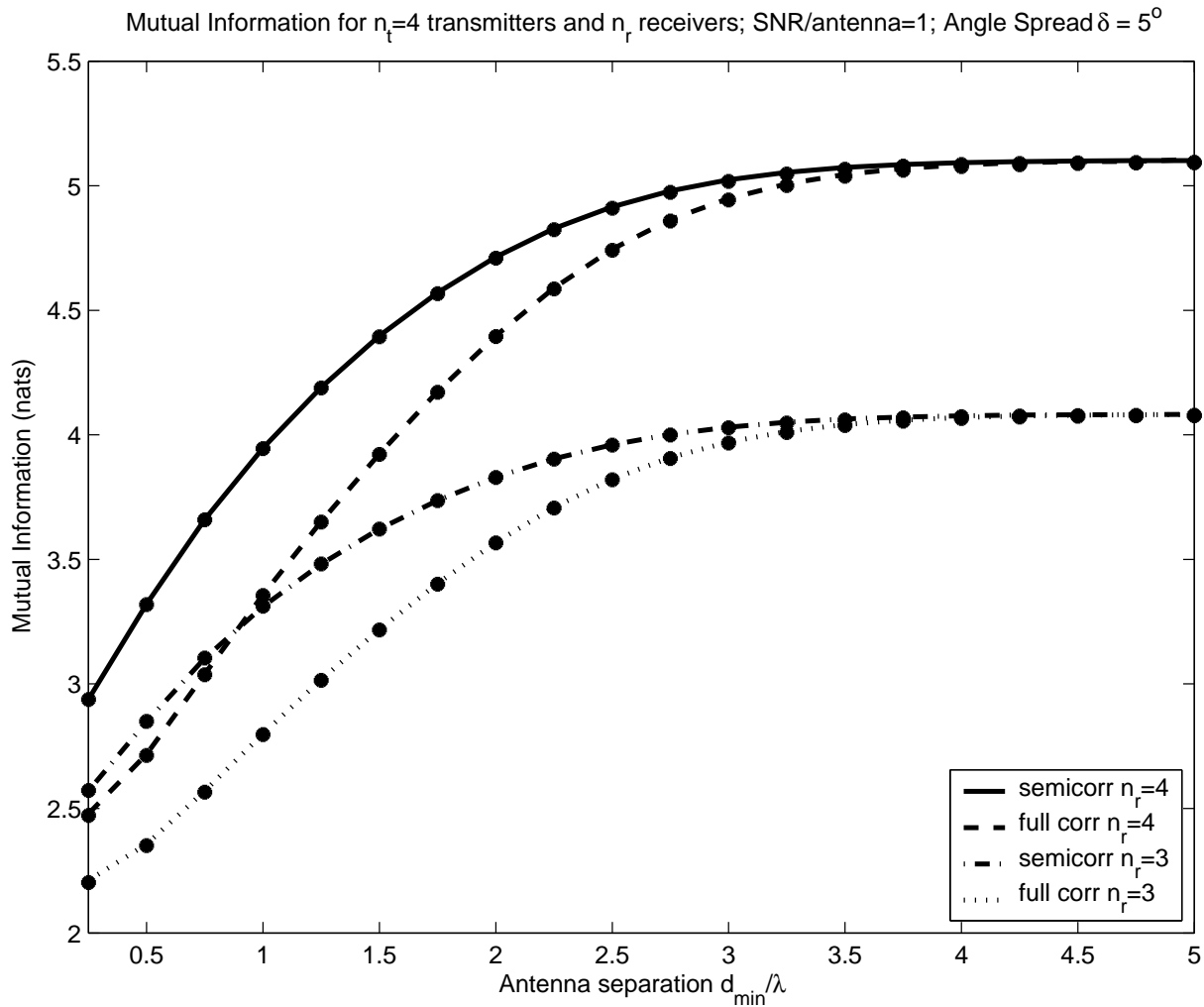}}
 \caption{Plot of mutual information for a
$\nt=4$-transmitting-antenna and a $\nr$-receiving-antenna array
pair for $\nr=3,4$. Both transmitter and receiver arrays are
uniformly spaced with $d_{min}$ spacing between neighboring
antennas. The solid and dashed curves have correlations between
elements on both antenna arrays given by
(\ref{eq:spatialcorrelations}) with angle spread $\delta=10^o$.
Once the eigenvalues of the correlation matrix are evaluated, they
are inserted in (\ref{eq:ergodic_cap}), which gives the ergodic
mutual information. The receiving array of the other two curves
(dotted and dot-dashed) has no antenna correlations. In this case
(\ref{eq:ergodic_cap_semicorr}) in Appendix \ref{app:Moment
generating function for semicorrelated channels} is used to
calculate the mutual information. We see that at large antenna
separation the antenna correlations become negligible and the
correlated cases approach the semicorrelated ones. The asterisks
represent the values of the simulated average capacities and have
been added as a cross-check. } \label{fig:cap_vs_dmin}
\end{figure}

\begin{figure}[htb]
\centerline{\epsfxsize=.9\columnwidth\epsffile{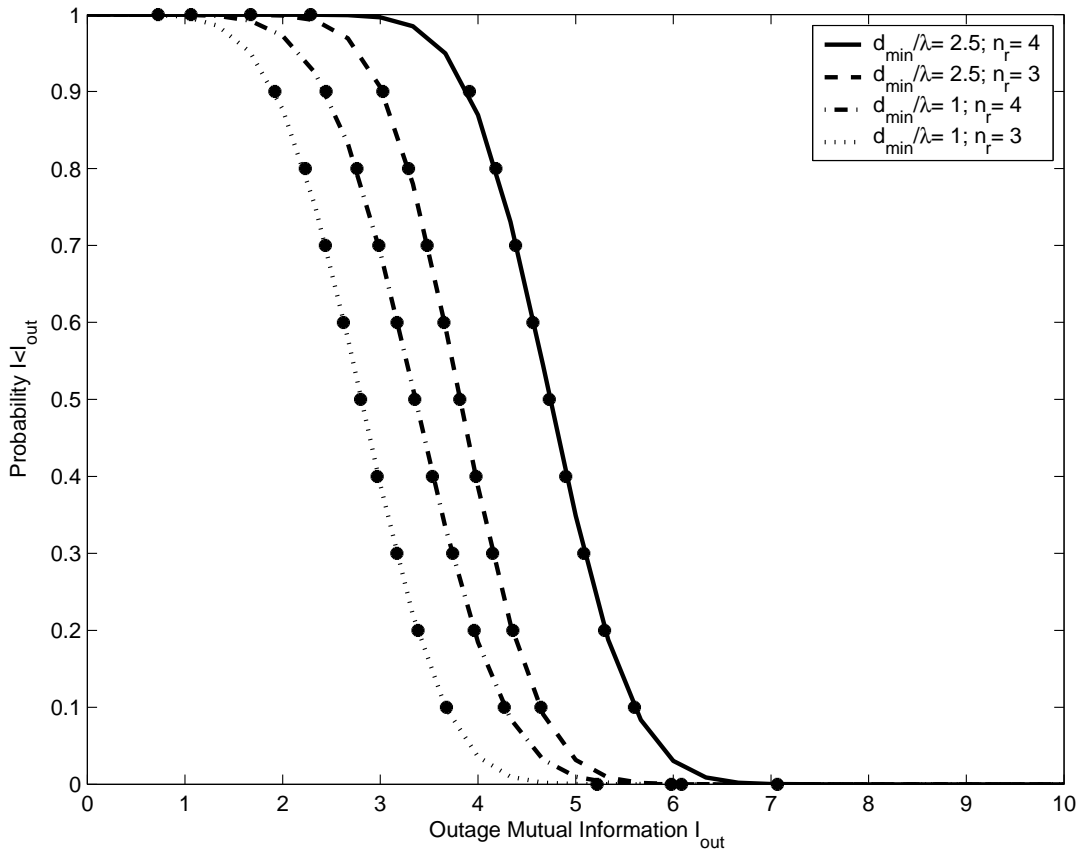}}
 \caption{Plot of the outage mutual
information for a $\nt=4$-transmitting-antenna and a
$\nr$-receiving-antenna array pair for $\nr=3,4$ and two different
array spacings $d_{min}$. The other parameters used are the same
as in Fig. \ref{fig:cap_vs_dmin}. The curves correspond to the
values of probability versus outage capacity, which were obtained
by numerically integrating (\ref{eq:outage_integral_def}). The
asterisks result from Monte-Carlo simulations and have been added
as verification of the numerical integral.} \label{fig:outage_cap}
\end{figure}

We will now apply the ergodic capacity equation
(\ref{eq:ergodic_cap}) derived above to evaluate the mutual
information for the a simple case of MIMO correlated channels.
Specifically, we will use the following simple model for the
dependence of the correlation coefficient $T_{ab}$ on the antenna
separation and angle spread. For concreteness, we assume that the
antennas form a uniform linear ideal antenna array with $
d_\lambda = d_{min}/\lambda$ the nearest neighbor antenna spacing
in wavelengths and we assume a Gaussian power azimuth spectrum
(with 2 dimensional propagation), i.e. the average incoming power
at the antenna array is $P(\theta) \propto
\exp[-(\theta/\delta)^2/2]$,\cite{Chizhik2000_CorrelatedMIMO1,Buehrer2000_CorrelationModel1}
 where $\delta$ is the angle spread in degrees measured from the
vertical to the array. This results in a $\vec T$ matrix with
elements
\begin{equation}
\label{eq:spatialcorrelations}
    T_{ab} = \int_{-180}^{180} \frac{d\phi}{\sqrt{2 \pi \delta^2}} e^{2
    \pi i (a - b) d_\lambda \sin(\phi\pi/180) - \phi^2/(2 \delta^2)}
\end{equation}
with $a,b= 1 \ldots \nt$ being the index of transmitting antennas.
Two situations will be considered: First, where the same
angle-spread appears in both transmitter and receiver, and second
where only the transmitter has the above correlations, with the
receiver having uncorrelated antennas. For simplicity, we assume
that the transmitted signal covariance is unity, that the noise is
uncorrelated and that the signal to noise ratio per transmitter
antenna is unity. In this case, one can evaluate the eigenvalues
of the correlation matrices and insert them in the appropriate
capacity equations. Fig. \ref{fig:cap_vs_dmin} shows the
comparison of the analytically obtained ergodic mutual information
for fully correlated and semicorrelated channels with simulations.
In Fig. \ref{fig:outage_cap} we plot the outage mutual information
obtained for a number of representative cases as calculated
numerically using (\ref{eq:outage_integral_def}) and compare them
with Monte Carlo simulations. It should be stressed however, that
results obtained analytically are exact, and we only include
Monte-Carlo simulations as a cross-check.

\section{Conclusion}
\label{sec:conclusion}

In conclusion, we have demonstrated a promising method recently
derived in the field of representations of Lie groups, which can
be used to calculate  complex integrals over unitary groups. We
applied this technique to recover in a simple way the joint
probability distribution of eigenvalues of the matrix
$\bG\bG^\dagger$, with $\bG$ being semicorrelated or having a
non-zero mean, which can then be used to calculate the moment
generating function of the mutual information for such Gaussian
MIMO channels. In addition, we used this method to calculate the
moment generating function of the mutual information for fully
correlated channels. From this moment generating function we
obtained the ergodic average of the mutual information and
evaluated the outage probability. We believe these methods are
general and applicable to a wide range of multi-antenna
communications problems. As an example, in Appendix
\ref{app:Application of Unitary Integrals on Unitary Space-Time
Encoding} we also analyzed their application  to unitary encoded
space-time transmission of MIMO systems, calculating, as a
particular example, the received signal distribution when the
channel matrix is correlated at the transmitter end. We expect
there will be further applications of these techniques in the
future.

\appendices

\section{Proof of Lemma \ref{lemma:Vandermonde determinant form of
d_r}}\label{app:Proof of Lemma Vandermonde determinant form of
d_r}

\begin{proof}
We start by writing $d_{\vec m}$ as
\begin{equation}\label{eq:d_r_vandermonde_form_proof1}
  d_{\vec m} = \left[\prod_{i=1}^M \frac{1}{(M-i)!}\right] \det\left[\frac{k_i!}{(k_i-M+j)!}\right]
\end{equation}
using (\ref{eq:vector_k_def_app}). The $i,j$-th element of the
matrix in the determinant can be expanded to
\begin{equation}\label{eq:expansion_of_det_lemma_vadermonde_form_d_m}
  k_i(k_i-1)\ldots(k_i-M+j+1)
\end{equation}
For example, the $M$th column has elements equal to unity, the
$(M-1)$th column has elements equal to $k_i$, etc. We note that
the product in
(\ref{eq:expansion_of_det_lemma_vadermonde_form_d_m}) in fact
vanishes when $k_i<M-j$, as noted after (\ref{eq:d_r_definition}).
As a result, the above expression is valid for every $i,j$ and
$k_i$. If we expand the above expression in powers of $k_i$ we get
\begin{equation}\label{eq:expansion_of_det_in_k_i_Lemma_vander_form_d_m}
  \sum_{p=1}^{M-j} a_{pj} k_i^{M-j-p+1}
\end{equation}
with the coefficient $a_{1j}=1$ for all $j=1,\ldots,M$ and all
other coefficients independent of $i$. We now subtract
$a_{M-j,j}k_i$, i.e. a multiple of column $M-1$ from all columns
$j=1,\ldots,(M-2)$. This operation leaves the value of the
determinant invariant, but reduces the upper limit in the
summation in
(\ref{eq:expansion_of_det_in_k_i_Lemma_vander_form_d_m}) to
$M-j-1$. We may now continue this process of subtracting off
$a_{p,j} k_i^{M-j-p+1}$ for $p=(M-j-1),(M-j-2),\ldots,2$ until the
only term appearing in the $ij$th term is $k_i^{M-j}$. This is
exactly the form of the Vandermonde matrix, with columns $j$ and
$M-j+1$ interchanged. Thus by reshuffling the columns to get the
usual ordering, and thereby picking up a term $(-1)^{M(M-1)/2}$,
we arrive to the form of (\ref{eq:d_r_vandermonde_form}).
\end{proof}

\section{}\label{app:Proof of Lemma: I_exponentially convergent}
In this appendix we will show that the quantity ${\cal I}(\vec T,
\vec R, \blambda)$ defined in (\ref{eq:P_lambda_i_fully_corr}) as
\begin{equation}\label{eq:I_def_in_lemma}
  {\cal I} = {\cal C}_M \Delta(\blambda) \int D\bU \int D\bV
\exp\left[-\Tr{ {\vec T^{-1} \bU  diag(\bmu) \bV^\dagger \vec
R^{-1} \bV  diag(\bmu)^\dagger \bU^\dagger}}\right]
\end{equation}
is exponentially small when any of the $\lambda$'s becomes
arbitrarily large. To prove this we start by showing that
\begin{lemma}\label{lemma:I_exponentially_convergent_firstlemma}
For any two $M$-dimensional positive definite matrices $\vec A$,
$\vec B$
\begin{equation}\label{eq:lemma:I_exponentially_convergent_firstlemma}
  \Tr{\vec A\vec B} \geq a_{min} \Tr{\vec B}
\end{equation}
where $a_{min}$ is
  the minimum eigenvalue of $\vec A$. Equality holds when $\vec A$
  is proportional to the unit matrix.
\end{lemma}
\begin{proof}
Let $a_1,\ldots,a_M$ be the eigenvalues of $\vec A$ and $\vec
U_{\vec A}$ be the unitary matrix diagonalizing $\vec A$, and
$b_1,\ldots,b_M$ and $\vec U_{\vec B}$ the corresponding
eigenvalues and unitary matrix for $\vec B$, i.e., $\vec A = \vec
U_{\vec A} diag(\vec a) \vec U_{\vec A}^\dagger$ and $\vec B =
\vec U_{\vec B} diag(\vec b) \vec U_{\vec B}^\dagger$. Then
\begin{eqnarray}\label{eq:proof_lemma_TrABgeq_aminTrB}
  \Tr{\vec A\vec B} &=& \Tr{diag(\vec a) \vec U_{\vec A}^\dagger \vec U_{\vec
  B} diag(\vec b^{1/2})) \left(\vec U_{\vec A}^\dagger \vec U_{\vec
  B} diag(\vec b^{1/2}))\right)^\dagger} \\ \nonumber
  &=& \sum_{i=1}^M a_i \sum_{j=1}^M\left|\left[\vec U_{\vec A}^\dagger \vec U_{\vec
  B} diag(\vec b^{1/2})\right]_{ij}\right|^2 \\ \nonumber
  &\geq& a_{min} \sum_{i=1}^M \sum_{j=1}^M \left|\left[\vec U_{\vec A}^\dagger \vec U_{\vec
  B} diag(\vec b^{1/2})\right]_{ij}\right|^2 \\ \nonumber
  &=& a_{min}\Tr{\vec U_{\vec A}^\dagger\vec B \vec U_{\vec A}} \\
  \nonumber
  &=& a_{min}\Tr{\vec B}
\end{eqnarray}
where $diag(\vec b^{1/2})$ is a diagonal matrix with diagonal
elements $b_i^{1/2}$, for $i=1,\ldots,M$. The inequality holds
since the term multiplying each $a_i$ is non-negative. One gets an
equality only if $\vec A$ is proportional to the unit matrix.
\end{proof}
We now apply this result to the trace in the exponent of
(\ref{eq:I_def_in_lemma})
\begin{eqnarray}\label{eq:proof_of_lemma_I_exp_convergent1}
\Tr{ \vec T^{-1} \bU  diag(\bmu) \bV^\dagger \vec R^{-1} \bV
 diag(\bmu)^\dagger \bU^\dagger} &\geq& t_{min} \Tr{ \bU  diag(\bmu) \bV^\dagger
\vec R^{-1} \bV  diag(\bmu)^\dagger \bU^\dagger}
\\ \nonumber %
&=& t_{min} \Tr{\vec R^{-1} \bV diag(\blambda) \bV^\dagger}
\\ \nonumber %
&\geq& t_{min} r_{min} \Tr{diag(\blambda)}
\end{eqnarray}
where $t_{min}$ and $r_{min}$ are the minimum eigenvalues of $\vec
T^{-1}$, $\vec R^{-1}$ respectively (or the inverses of the
maximum eigenvalues of $\vec T$, $\vec R$, which are assumed
non-zero). As a result,
\begin{eqnarray}\label{eq:proof_of_lemma_I_exp_convergent2}
{\cal I}_{\vec T,\vec R, \blambda} &\leq& {\cal C}_M
\Delta(\blambda) \int D\bU \int D\bV \exp\left[-t_{min} r_{min}
\Tr{diag(\blambda)}\right]\\ \nonumber %
&=& {\cal C}_M \Delta(\blambda) \exp\left[-t_{min} r_{min}
\sum_{i=1}^M \lambda_i\right]
\end{eqnarray}
Since $t_{min}r_{min}>0$ we finally have
\begin{lemma}\label{lemma:I_exponentially convergent}
The quantity ${\cal I}(\vec T, \vec R, \blambda)$ in
(\ref{eq:I_def_in_lemma}) is bounded by an exponential function of
any $\lambda_i$ when for any $i=1,\ldots,M$, $\lambda_i$ becomes
arbitrarily large.
\end{lemma}

\section{Limit of Ratios of Functions}
\label{app:Proof of Lemma Vandermonde_finite_x0}

This appendix concerns limits of ratios of the form
\begin{equation}\label{eq:R_function_app_vandermonde_lemma}
  R([x_1,\ldots\, x_M]) =   \frac{\det[f_i(x_j)]}{\Delta([x_1,\ldots\,x_M])}
\end{equation}
when several $x_j$'s become equal. In the above equation, $i,j=1,
\ldots, M$, $\Delta(\vec x)$ is the Vandermonde determinant and
$\det[f_i(x_j)]$ is the determinant of an $M\times M$ matrix whose
$i,j$-th element is $f_i(x_j)$.

\begin{lemma}\label{lemma:vandermonde_finite_x0}
When two or more of the $x_j$'s in
(\ref{eq:R_function_app_vandermonde_lemma}) become equal, then
both numerator and denominator are zero. However, the ratio is
generally well defined. In particular,
\begin{eqnarray}\label{eq:lemma_vandermonde_finite_x0}
& &     \lim_{x_1, x_2, \ldots,\, x_p \rightarrow x_0} R([x_1,
\ldots
\, x_M]) = \\
   & &  \frac{\det\left[ f_i(x_0)  ;  f_i^{(1)}(x_0) ;
    \ldots  ; f_i^{(p-1)}(x_0) ;  f_i(x_{p+1})  ;
    f_i(x_{p+2}) ; \ldots ;  f_i(x_M) \right]}{ \Delta(x_{p+1} \ldots x_M)
    \prod_{i={p+1}}^M  (x_i-x_0)^p \,\, \prod_{j=1}^{p-1} j!}  \nonumber
\end{eqnarray}
where $f^{(k)}$ represents the $k^{th}$ derivative of the function
$f$.  (For clarity, we separate the columns by semicolons, and
note that the different rows of the determinant correspond to
different values of $i = 1, \ldots, M$).
\end{lemma}

\begin{proof}
Proof will be by induction on $p$.  The theorem is trivial for
$p=1$. Suppose the theorem holds up to some given $p$.  Then we
would like to take the limit as $x_{p+1} \rightarrow x_0$.
Examining the matrix of the numerator, we write the elements of
the $p+1$'st column in a Taylor expansion
\begin{equation}
    f_i(x_{p+1}) =  \sum_{j=0}^\infty \frac{f_i^{(j)}(x_0)}{j!} (x_{p+1}-x_0)^j
 \end{equation}
We will now use the fact that you can subtract any multiple of one
column from another column without changing the value of the
determinant.  Since the $j^{th}$ column is $f_i^{(j)}(x_0)$ we can
thus subtract off the first $p$ terms of this sum $(j=0 \ldots
p-1)$ without changing the determinant, leaving us with the
$p+1$'st column given by
\begin{equation}
\label{eq:lemma_vandermonde_numerator_expansion} f_i(x_{p+1})
\rightarrow  f_i^{(p)}(x_0) \frac{(x_{p+1}-x_0)^{p}}{p!} +
\mbox{higher powers of } (x_{p+1}-x_0)
\end{equation}
We now examine the denominator in
(\ref{eq:lemma_vandermonde_finite_x0}).  Again, we assumed that
the theorem holds for some given $p$.  Thus, as we take $x_{p+1}
\rightarrow x_0$ the denominator becomes
\begin{eqnarray} \label{eq:lemma_vandermonde_denominator_expansion}
& & \lim_{x_{p+1} \rightarrow x_0}  \Delta(x_{p+1} \ldots x_M)
    \prod_{i={p+1}}^M  (x_i-x_0)^p \,\, \prod_{j=1}^{p-1} j! \,\,\,\, =  \\
    & &  (x_{p+1}-x_0)^p \left[ \prod_{i={p+2}}^M (x_i - x_{p+1}) \right]  \Delta(x_{p+2} \ldots x_M)
    \prod_{i={p+2}}^M  (x_i-x_0)^p \,\, \prod_{j=1}^{p-1} j!     \nonumber
\end{eqnarray}
Plugging in the results of
(\ref{eq:lemma_vandermonde_denominator_expansion}) and
(\ref{eq:lemma_vandermonde_numerator_expansion}) into
(\ref{eq:lemma_vandermonde_finite_x0}) proves the lemma for $p+1$.
\end{proof}

A related result can be shown for the case when some elements of
the vector $\vec x$ go to infinity:

\begin{lemma}
\label{lemma:vandermonde_infinite_x} If the asymptotic behavior of
$f_i(x_j)$, when $x_j\rightarrow \infty$ is of the form
\begin{equation}\label{eq:vandermonde_lemma_asymptotic_f_i_x_j_form}
  f_i(x_j) \approx x_j^{M-1} \sum_{k=0}^\infty \hat{f}_i^{(k)}
  x_j^{-k}
\end{equation}
then we have the following limit
\begin{eqnarray}\label{eq:vanderlemma_main}
& &     \lim_{x_1, x_2, \ldots,\, x_p \rightarrow \infty}
R(x_1,\ldots\, x_M) = \\
   & &  \frac{\det\left[ \hat{f}_i^{(0)}  ;  \hat{f}_i^{(1)} ;
    \ldots  ; \hat{f}_i^{(k)} ;  f_i(x_{p+1})  ;
    f_i(x_{p+2}) ; \ldots ;  f_i(x_M) \right]}{ \Delta(x_{p+1} \ldots
    x_M)} \nonumber
\end{eqnarray}
\end{lemma}

\section{Moment generating function for semicorrelated and
non-zero mean channels} \label{app:Moment generating function for
semicorrelated and non-zero mean channels}

In this appendix we use the joint probability distribution
$P(\{\lambda_i\})$ derived in Sections \ref{sec:Case of
semi-correlated G with zero mean} and \ref{sec:Case of
uncorrelated G with non-zero mean G0} to calculate the
corresponding moment generating function $g(z)$,
(\ref{eq:g_z_lambda}). We start with the following Lemma.
\begin{lemma}
\label{lemma:unzipping_determinants} Let $F(\{\lambda_i\})$ be a
function of $\lambda_i$, for $i=1,\ldots,N$, defined as
\begin{equation}\label{eq:app_gen_function_lemma_F_lambda_i}
  F(\{\lambda_i\}) = {\cal K} \prod_{i=1}^N f(\lambda_i) \Delta(\blambda)
  \det \vec P(\blambda)
\end{equation}
where ${\cal K}$ is a constant independent of $\lambda_i$ and the
$M\times M$ matrix $\vec P$, with $M\geq N$, has elements
\begin{equation}\label{eq:app_gen_function_lemma_det_F_lambda_i}
  P_{ij} = \left\{
\begin{array}{cc}
  p_i(\lambda_j) & j\leq N   \\
  q_{ij}         & j>N
  \end{array}
\right.
\end{equation}
for some function $p_i(x)$. Then integrating $F$ over $\lambda$ we
have
\begin{equation}\label{eq:app_gen_function_lemma_int_F_lambda_i}
  \prod_{j=1}^N \int_0^\infty d\lambda_j F(\{\lambda_i\}) = N! \det  \tilde{\vec P}
\end{equation}
where the $M\times M$ matrix $\tilde{\vec P}$ has elements
\begin{equation}\label{eq:app_gen_function_lemma_det_P_tilde_ij}
  \tilde{P}_{ij} = \left\{
\begin{array}{cc}
  \int_0^\infty d\lambda f(\lambda) \lambda^{i-1} p_j(\lambda) & i\leq N   \\
  q_{ij}         & i>N
  \end{array}
\right.
\end{equation}
\end{lemma}
\begin{proof}
We expand the two determinants as in (\ref{eq:det_expansion_def})
resulting in
\begin{eqnarray}\label{eq:app_gen_function_lemma_proof_int_F_lambda_i}
  \prod_{j=1}^N \int_0^\infty d\lambda_j F(\{\lambda_i\}) &=&
  \sum_{\vec a}\sum_{\vec b} \sgn(\vec a)\sgn(\vec b)
  \prod_{i=1}^N \int_0^\infty d\lambda_i f(\lambda_i) \lambda_i^{a_i-1} p_{b_i}(\lambda_i)
  \prod_{i=N+1}^M q_{i b_i} \\ \nonumber %
  &=& \sum_{\vec a}\sum_{\vec b} \sgn(\vec a)\sgn(\vec b)
  \prod_{i=1}^N \tilde{P}_{a_i b_i} \prod_{i=N+1}^M q_{i b_i}\\ \nonumber %
  &=& N! \sum_{\vec b} \sgn(\vec b)
  \prod_{i=1}^M \tilde{P}_{i b_i} \\ \nonumber %
  &=& N! \det \tilde{\vec P}
\end{eqnarray}
\end{proof}
We now apply this lemma to the two cases of semicorrelated and
non-zero mean channels analyzed in Sections \ref{sec:Case of
semi-correlated G with zero mean} and \ref{sec:Case of
uncorrelated G with non-zero mean G0}.

\subsection{Moment generating function for semicorrelated
channels}\label{app:Moment generating function for semicorrelated
channels}

We can immediately apply the above lemma to calculate the moment
generating function. We distinguish two cases:

\subsubsection{$N=\nt\leq\nr=M$} In this case, the form of the matrix $\vec P$
simplifies, since there are no $q_{ij}$ terms (see
(\ref{eq:P_lambda_i_semicorr_nt_less_nr})) and we thus have
$f(\lambda)=(1+\lambda)^z\lambda^{\nr-\nt}$ and $p_j(\lambda)=
e^{-t_j\lambda}$, resulting in
\begin{equation}\label{eq:app_gen_function_semi_corr_nt_less_nr}
  g(z) =  (-1)^{\nt(\nt-1)/2}\prod_{j=1}^\nt t_j^\nr \frac{\det \vec L}{\Delta(\vec t)}
\end{equation}
with the $\nt\times \nt$ matrix $\vec L$ has elements
\begin{equation}\label{eq:app_gen_function_semicorr_L_ij_nt_less_nr}
L_{ij} = \Psi(\nr-\nt+j,\nr-\nt+j+z+1,t_i)
\end{equation}
where $\Psi(a,b,z)$ is the confluent hypergeometric function
\cite{Gradshteyn_Ryzhik_book}, given by
\begin{equation}\label{eq:confluent_hyperg_def}
  \Psi(a,b,z) = \frac{1}{\Gamma(a)} \int_0^\infty dt e^{-tz} t^{a-1}
  (1+t)^{b-a-1}
\end{equation}
Note  we absorbed the product of factorials in
(\ref{eq:P_lambda_i_semicorr_nt_less_nr}) into the $\Psi$
function.

\subsubsection{$M=\nt>\nr=N$} In this case we use (\ref{eq:P_lambda_i_semicorr_nt_greater_nr}) and
set $f(\lambda)=(1+\lambda)^z$, $q_{ij} = t_i^{j-\nr-1}$ and
$p_i(\lambda)= e^{-t_i\lambda}$, resulting in the following
expression for $g(z)$
\begin{equation}\label{eq:app_gen_function_semi_corr_nt_great_nr}
  g(z) =  (-1)^{\nt(\nt-1)/2}\prod_{i=1}^\nt t_i^\nr \frac{\det \vec L}{\Delta(\vec t)}
\end{equation}
with the $\nt\times \nt$ matrix $\vec L$ has elements
\begin{equation}\label{eq:app_gen_function_semicorr_L_ij_nt_greater_nr}
L_{ij} = \left\{
\begin{array}{cc}
  t_i^{j-1}         & 1\leq j\leq\nt-\nr   \\
  \Psi(\nr-\nt+j,\nr-\nt+j+z+1,t_i) & j>\nt-\nr
  \end{array}
\right.
\end{equation}

\subsubsection{Ergodic Capacity}

By employing methods similar to Section \ref{sec:Ergodic Capacity
for fully correlated channels} the expression for the ergodic
mutual information of semicorrelated channels, which is analogous
to (\ref{eq:ergodic_cap}) and in agreement with
\cite{Chiani2003_CapacityOfSpatiallyCorrelatedMIMOChannels,
Smith2003_CapacityMIMOsystems_with_semicorrelated_flat_fading}
\begin{equation}\label{eq:ergodic_cap_semicorr}
  E[I] = \sum_{j=1}^N \frac{\det \tilde{\vec L}_j}{\det\vec L}
\end{equation}
where $\vec L$ is given by
(\ref{eq:app_gen_function_semi_corr_nt_less_nr}) or
(\ref{eq:app_gen_function_semi_corr_nt_great_nr}) evaluated at
$z=0$ and $\tilde{\vec L}_j$ is given by the matrix $\vec L$, with
the elements of the column $j+\max(0,\nt-\nr)$ substituted by
\begin{eqnarray}\label{eq:D_ij_semi_corr_def}
  D_{ij} &=& \frac{(-1)^{M-\nt+j}}{(M-\nt+j-1)!} \left[\frac{e^{t_i} Ei(-t_i)}{t_i}\right]^{(M-\nt+j-1)}
\end{eqnarray}
with the superscript in the last term indicating the number of
derivatives applied.

\subsection{Moment generating function for channels with non-zero mean}
\label{app:Moment generating function for channels with non-zero
mean}

Applying Lemma \ref{lemma:unzipping_determinants} to
(\ref{eq:P_lambda_non_zero_mean_fin2}) we obtain the following
expression for $g(z)$
\begin{equation}\label{eq:app_gen_function_nz_mean}
g(z)= (-1)^{(N_0+N)(M-1)}
\frac{\prod_{i=1}^{M-N}\left[\frac{(N+i-1)!}{(i-1)!}\right]}{
\prod_{j=1}^{M-N_0}\left[(j-1)!\right]^2 } \prod_{j=1}^{N_0}
\left[\gamma_j^{N_0-M} e^{-\gamma_j}\right] \frac{\det\vec
L}{\Delta(\bgamma)}
\end{equation}
$\vec L$ is a $M\times M$ matrix with elements
\begin{equation}\label{eq:nzmean_g_z_Lmatrix}
  L_{ij} = \left\{
\begin{array}{cc}
  \int_0^\infty d\lambda \lambda^{j-1} (1+\lambda)^z e^{-\lambda}I_0(2\sqrt{\gamma_i\lambda}) & i\leq N_0, j\leq N   \\
  (i+j-N_0-1)!\Psi(i+j-N_0-1,i+j-N_0+z,1) & i> N_0, j\leq N
  \\
  \gamma_i^{j-N-1} & i\leq N_0, j>N  \\
  \left[(i-N_0-1)!\right]^2\delta_{i-N_0,j-N} & i> N_0, j> N
\end{array}
\right.
\end{equation}

\section{Other directions: Application of Unitary Integrals to Unitary Space-Time
Encoding} \label{app:Application of Unitary Integrals on Unitary
Space-Time Encoding}

In this section we will briefly show another application of
integration over unitary groups, where the character expansion and
other results analyzed above may be applied, namely in the field
of unitary space-time encoding. This is a scheme suggested
recently by \cite{Marzetta1999_USTM, Hochwald2000_USTM,
Hassibi2002_USTM, Hughes2000_DUSTM}, in which the signal is
encoded using unitary matrices across multiple transmit antennas
and time-slots. The interest in this encoding method lies in its
very good performance for large signal to noise ratios in the
absence of any channel information at the
receiver.\cite{Hassibi2002_USTM}

In particular, we assume that the channel matrix $\bG$ is constant
over $T$ time-intervals and then changes completely. In this case
the channel equation (\ref{eq:basic_channel_eq}) needs to be
modified to\cite{Marzetta1999_USTM}
\begin{equation}\label{eq:basic_channel_eq_ustm}
  \vec Y = \bG \vec X + \vec Z
\end{equation}
where $\bG$ is a $\nt\times \nr$ matrix with the channel
coefficients from the transmitting to the receiving arrays, $\vec
Z$ is the $\nr \times T$ additive noise matrix, while $\vec Y$ and
$\vec X$ are $\nr\times T$ and $\nt \times T$ dimensional matrices
representing the output and input signals. The noise elements
$Z_{ij}$ are assumed to be Gaussian i.i.d. with unit variance,
while $\bG$ is semicorrelated at the transmitter end, i.e.
$E[G_{ia} G_{jb}] = \delta_{ij} T_{ab}$. Their instantaneous
values are assumed to be unknown to both the transmitter and the
receiver.

Recently \cite{Hassibi2002_USTM} analyzed the mutual information
of this problem, when the input distribution $\vec X$ is unitary,
i.e. when
\begin{equation}\label{eq:Phiunitary}
 \vec X \vec X^\dagger = \vec I_\nt
\end{equation}
(but $\vec X^\dagger \vec X\neq \vec I_T$). One can thus express
$\vec X$ as
\begin{equation}\label{eq:X=UI_M}
  \vec X = \vec J_\nt \vec U
\end{equation}
where $\vec U$ is an $T\times T$ unitary matrix and $\vec J_\nt$
is a $\nt\times T$ matrix with $J_{ab}=\delta_{ab}$.
(\ref{eq:X=UI_M}) includes only $\nt$ out of $T$ orthogonal
$T$-dimensional vectors. To obtain the capacity of such encoding
structures, it is essential to first calculate the received signal
probability distribution, which involves an integration over all
transmitted signals, which in this case are unitary matrices.

Since both $\bG$ and $\vec Z$ are Gaussian matrices, the
conditional distribution of $\vec Y$ on $\vec X$, $p(\vec Y|\vec
X)$ will also be Gaussian, with density
\begin{eqnarray}\label{eq:p(X|Phi)_def}
  p(\vec Y|\vec X) &=&     \frac{\exp\left(-\Tr{ \vec Y
 \left[\vec I_{T} + \vec X^\dagger \vec T\vec X \right]^{-1}
  \vec Y^\dagger}\right)}{\pi^{T\nr}\det(\vec I_T+\vec X^\dagger \vec T\vec X)^{\nr}}
  \nonumber \\
  &=&
\frac{\exp\left(-\Tr{ \vec Y
 \left[\vec I_{T} - \vec X^\dagger \frac{\vec T}{\vec I_\nt+\vec T}\vec X \right]
  \vec Y^\dagger}\right)}{\pi^{T\nr}\det(\vec I_\nt+\vec T)^{\nr}}
\end{eqnarray}
where the second equation results from $\vec X$ being unitary,
(\ref{eq:Phiunitary}). The received signal distribution is
\begin{equation}\label{eq:p(X)}
  p(\vec Y) = \int d\vec X \, p(\vec  Y|\vec X)
\end{equation}
\cite{Hassibi2002_USTM} obtained a closed form expression for the
case $T_{ab} = \delta_{ab}$ using different methods. Here we will
obtain a closed form expression for a general positive definite
$\vec T$.

We can rewrite the above integral over $\vec X$ using
(\ref{eq:X=UI_M}) as
\begin{eqnarray}\label{eq:p(Y)int}
  p(\vec Y) =  \int D\vec U
\frac{\exp\left(-\Tr{ \vec Y
 \left[\vec I_{T} - \vec U^\dagger \tilde{\vec T} \vec U \right]
  \vec Y^\dagger}\right)}{\pi^{T\nr}\det(\vec I_\nt+\vec T)^{\nr}}
\end{eqnarray}
where
\begin{equation}\label{eq:T_tilde_def}
  \tilde{\vec T} = \vec J_\nt \frac{\vec T}{\vec I_\nt + \vec T}
  \vec J_\nt^\dagger
\end{equation}
We immediately see that the integral in (\ref{eq:p(Y)int}) is of
the form of (\ref{eq:P_lambda_i_semicorr_int}), which allows us to
immediately obtain the result. We are interested in the case
$T\geq \nt$.\cite{Hassibi2002_USTM} Defining the $T$-dimensional
vector $\tilde{\vec t}$ so that $\tilde{t}_i=t_i=T_i/(1+T_i)$ for
$i=1,\ldots,\nt$, where $T_i$ are the eigenvalues of $\vec T$, and
using (\ref{eq:P_lambda_i_semicorr_MN_gen}) we have
\begin{eqnarray}
\label{eq:p(Y)_int result1} %
p(\vec Y)= \frac{\prod_{p=1}^{T-1}p!}{\prod_{j=1}^\nt (1+T_j)}
\frac{\det\left(e^{y_i \tilde{t}_j}\right)}{\Delta(\vec
y)\Delta(\tilde{\vec t})}\exp(-\sum_i y_i)
\end{eqnarray}
where $\vec y$ is the $T$-dimensional vector of eigenvalues of
$\vec Y^\dagger\vec Y$. We now need to take
$\tilde{t}_i\rightarrow 0$ for $i=\nt+1,\ldots, T$. Also, if
$T>\nr$ we need to set $y_i=0$ for $i=\nr+1,\ldots, T$. These
limits can be taken by using Lemma
\ref{lemma:vandermonde_finite_x0}. The final result is
\begin{eqnarray}
\label{eq:p(Y)_int result2} %
p(\vec Y)= \frac{\prod_{p=T-\nt}^{T-1}p!}{\prod_{j=1}^\nt
(1+T_j)\prod_{i=1}^{T-Q-1}i!} \frac{\det\vec L}{\Delta(\vec
y)\prod_{i=1}^Qy_i^{T-Q}\Delta(\vec t)\prod_{j=1}^\nt t_i^{T-\nt}}
\exp(-\sum_{i=1}^Q y_i)
\end{eqnarray}
where $Q=\min(T,\nr)$ and the $T$-dimensional matrix $\vec L$ has
elements
\begin{equation}\label{eq:p(Y)_final_result_matrix L}
  L_{ij} = \left\{
\begin{array}{cc}
  e^{y_it_j} & i\leq Q, j\leq \nt   \\
  y_i^{j-\nt-1} & i\leq Q, j> \nt  \\
  t_j^{i-Q-1} & i>Q, j\leq \nt  \\
  (i-\nr-1)!\delta_{i-\nr,j-\nt} & i> Q, j> \nt
\end{array}
\right.
\end{equation}



\end{document}